# Evaluating GPT- and Reasoning-based Large Language Models on Physics Olympiad Problems: Surpassing Human Performance and Implications for Educational Assessment

Paul Tschisgale[1], Holger Maus[1], Fabian Kieser[2], Ben Kroehs[3], Stefan Petersen[1], Peter Wulff[4]

[1] Department of Physics Education, Leibniz Institute for Science and Mathematics Education, Kiel, Germany

[2] Department of Physics-Physics Education Research, Free University of Berlin, Berlin, Germany

[3] Kirchhoff Institute for Physics, Ruprecht Karl University of Heidelberg, Heidelberg, Germany

[4] Department of Physics and Physics Education Research, Ludwigsburg University of Education, Ludwigsburg, Germany

**ABSTRACT**

Large language models (LLMs) are now widely accessible, reaching learners at all educational levels. This development has raised concerns that their use may circumvent essential learning processes and compromise the integrity of established assessment formats. In physics education, where problem solving plays a central role in instruction and assessment, it is therefore essential to understand the physics-specific problem-solving capabilities of LLMs. Such understanding is key to informing responsible and pedagogically sound approaches to integrating LLMs into instruction and assessment. This study therefore compares the problem-solving performance of a general-purpose LLM (*GPT-4o*, using varying prompting techniques) and a reasoning-optimized model (*o1-preview*) with that of participants of the German Physics Olympiad, based on a set of well-defined Olympiad problems. In addition to evaluating the correctness of the generated solutions, the study analyzes characteristic strengths and limitations of LLM-generated solutions. The results of this study indicate that both tested LLMs (*GPT-4o* and *o1-preview*) demonstrate advanced problem-solving capabilities on Olympiad-type physics problems, on average outperforming the human participants. Prompting techniques had little effect on *GPT-4o*'s performance, and *o1-preview* almost consistently outperformed both *GPT-4o* and the human benchmark. The main implications of these findings are twofold: LLMs pose a challenge for summative assessment in unsupervised settings, as they can solve advanced physics problems at a level exceeding that of top-performing students, making it difficult to ensure the authenticity of student work. At the same time, their problem-solving capabilities offer potential for formative assessment, where LLMs can support students in evaluating their own problem solutions.

# I. INTRODUCTION

Large language models (LLMs) are now widely accessible to the public, including students at all educational levels. Since the release of ChatGPT in November 2022, there has been rapid progress in LLM development, accompanied by striking improvements in their apparent capabilities. In parallel, concerns have been raised about students using these tools to circumvent meaningful learning processes [1,2] and to undermine the integrity of unsupervised exams or written assignments [3,4]. More broadly, LLMs have become central to ongoing discussions about the role of artificial intelligence (AI) in education (e.g., [5–8]). It is therefore essential for educators and educational researchers to remain informed about the rapid advancements in AI—particularly with regard to LLMs—in order to investigate how these technologies can be integrated into education in effective and responsible ways.

A step in this direction is to examine how LLMs relate to specific forms of instruction and assessments that are central in different disciplines. In physics education, problem solving is central to instruction and assessment, as it relates to both conceptual understanding and the application of physics-specific knowledge in structured and goal-oriented ways. Physics problem-solving abilities are pivotal to master for students planning to engage in a physics-related career [9,10]. Supporting the development of this ability is therefore a central goal of physics education and physics education research [11,12]. Understanding the apparent capabilities of LLMs in this context is important, particularly in light of emerging evidence connecting their problem-solving and assessment capabilities [13].

LLMs have shown promising performance on conceptual physics questions and physics textbook problems (e.g., [14,15]), generating interest in their potential as tools to support learning. Further research examined more advanced physics problems (e.g., [16–18]), however, often a focus was placed on final answer correctness, offering limited insight into LLMs' problem-solving processes and how they compare to problem-solving processes of actual students—including the extent to which LLMs display human-like misconceptions. At the same time, the effectiveness of specific prompting techniques—that is, phrasing instructions in particular ways to elicit better responses from LLMs—remains inconclusive (e.g., [16,19]). Moreover, the latest generation of LLMs explicitly optimized for reasoning (e.g., OpenAI's *o1* model) remains largely unexplored in research, despite claims that they offer substantially improved problem-solving capabilities compared to earlier models [20]. Overall, there is a need to move beyond evaluations based solely on final answer correctness and to examine entire solution processes—particularly in relation to the role of different prompting techniques and the yet-untapped potential of the latest generation of reasoning-optimized LLMs.

To address this gap, the present study systematically compares the performances of a general-purpose GPT model (*GPT-4o*; using different prompting techniques), a reasoning-optimized model (*o1-preview*), and students participating in the German Physics Olympiad on a set of advanced problems taken from the German Physics Olympiad [21]. These problems, characterized by conceptual richness and multi-step reasoning demands, serve as a rigorous benchmark for assessing LLMs' problem-solving capabilities beyond routine textbook problems that were likely seen during LLM training. In addition to evaluating final answer correctness and intermediate reasoning steps, the study examines LLMs' strengths and weaknesses in detail. Findings from this study thereby contribute to the growing discourse on the nature and depth of LLMs' apparent physics understanding and physics-related capabilities (e.g., [22,23]). Moreover, the results offer implications for educational assessment in the age of advanced LLM systems, offering concrete insights for rethinking assessment practices in physics education.

## II. BACKGROUND

### A. Physics problem solving

We consider a physics problem to be any task that requires physics-specific analysis and reasoning, progressing from an initial problem situation towards a goal state [24]. Problem solving then refers to the process of successively transforming this initial state, as determined by a specific problem situation, toward the goal state [25]. Problems can generally be categorized along a continuum ranging from well-defined to ill-defined problems [26]. Well-defined problems, such as those commonly found in textbooks, are characterized by a clear problem statement, a precisely defined goal state, and a well-specified set of actions required to reach a solution [27]. In contrast, problems encountered in real-world physics professions tend to be more ill-defined, as they often lack a clearly defined initial state or a specific goal state [28]. Unlike textbook problems, real-world problems do not always have a single correct or optimal solution, leading to multiple possible solution paths.

#### 1. Problem-solving processes

The process of solving well-defined problems has been extensively investigated, leading to the development of both domain-general (e.g., [29]) and domain-specific problem-solving process models (for physics, see e.g., Ref. [30]). These models generally outline four sequential phases of problem solving: (i) representing the problem from a physics perspective, (ii) selecting an appropriate solution strategy, (iii) executing the strategy, and (iv) evaluating the final solution (e.g., [31–33]).

Among these phases, the initial two—problem representation and strategy selection—are generally considered the most crucial for successful problem solving [32,34]. Effective problem representation involves modeling the situation from a physics perspective, identifying relevant principles, and constructing a coherent (mental) model. Based on this, a (potentially expedient) strategy is selected which involves asking which concepts can be applied, how they can be applied, and under which conditions [35]. Once a suitable problem representation and strategy are in place, the strategy execution phase typically involves more routine mathematical procedures that are relatively independent of the underlying physics. Finally, the evaluation of the solution should entail checking the plausibility or consistency of the final result, or even better of all steps leading to the final result.

While these models provide valuable frameworks for teaching metacognitive strategies and guiding structured problem solving [36–38], they also represent idealized simplifications of how students actually solve problems [39,40]. In practice, the outlined phases may occur in varying sequences and may be revisited multiple times within a single problem-solving process [41]. Particularly, there also exists differences between high- and low-performing problem solvers (e.g., [41–45]). Yet, to our knowledge, no study has systematically investigated whether and how this idealized problem-solving process structure is reflected in solutions to physics problems generated by large language models, or to what extent prompting a large language model to follow such a structure improves its problem-solving performance.

**2. High-level problem solving in the Physics Olympiad**

Problem solving also plays a central role in the German Physics Olympiad, in which secondary school students engage with both theoretical and experimental physics problems across four competition stages [21]. While the entry stage involves solving these problems as homework over an extended period, written examinations are used from the second stage onward. These problems are designed not only to test students' conceptual physics knowledge but particularly to assess their physics problem-solving ability, i.e., their ability to apply this conceptual knowledge in diverse and challenging problem-solving contexts. This ability was particularly shown to be the main predictor of success in the Olympiad [46]. Olympiad problems are generally well-defined, meaning all necessary information for solving them is provided in the problem description, and possible solution paths are typically limited. However, unlike most textbook problems, they tend to be more complex, requiring participants to construct sophisticated problem representations and integrate multiple physics concepts to reach a solution, i.e., multi-step reasoning is generally required. In this regard, these problems usually involve the application of mathematics [47]. Efforts are also made to design problems that are innovative in the sense that their solutions are not readily accessible through a simple web search or by consulting a few physics textbooks—though this may not hold true for all problems. Moreover, many of the problems are not publicly shared through the Internet, and thus likely not part of the Common Crawl of the Internet, which is part of the training data for LLMs. The average difficulty of the problems increases across the competition stages, and their scope broadens to include more advanced topics: while first-stage problems typically align with standard school curricula, problems in subsequent stages adhere to the International Physics Olympiad syllabus [48] which also encompasses advanced physics topics extending beyond typical school instruction in Germany.

**B. Large language models**

*1. Basic functionality*

Large language models (LLMs) are a class of generative AI designed to interpret and generate natural language. While their primary strength lies in natural language understanding and generation, their capabilities now extend to domains such as programming, mathematics, and logic [49–51]. These models are typically trained in two stages: large-scale pretraining on massive and diverse text corpora (drawn from sources such as books, articles, Wikipedia, and the Common Crawl of the Internet), followed by fine-tuning (often supervised or reinforced) to improve performance on specific tasks and align outputs with human expectations [52].

At their core, the majority of LLMs operates via autoregressive inference: Given an input prompt (typically a user-provided question or instruction), they generate text—one token[1] at a time—by sampling the next token from a probability distribution over the model's vocabulary, conditioned on the input prompt and all previously generated tokens [53]. This conditional probability distribution reflects the model's estimate of which tokens are most likely to come next, based on statistical patterns between words learned during training. This process continues iteratively until a stopping criterion is met (e.g., reaching an end-of-sequence token or a predefined maximum length). The next-token generation is influenced both by how the input prompt is phrased and by certain model parameters. One of the most important model parameters is *temperature*, which controls the randomness of token generation. Lower temperature values concentrate probability mass on the most likely tokens, leading to more deterministic and focused outputs. Higher temperature values flatten the distribution, encouraging more diverse and creative token completions, but also increasing the risk of incoherence or off-topic content. For readers interested in a more in-depth—though still accessible—explanation of how LLMs work, we refer to Ref. [54] or Ref. [55].

---

[1] A token is a unit of text used by language models during processing and generation. Depending on the model's tokenization scheme, a token may correspond to a word, subword, or even a single character.

## 2. GPT vs. reasoning models: Two modes of AI thinking

Among the most influential LLMs are those developed by OpenAI. Since September 2024, OpenAI offers two types of models: Generative pre-trained transformer or GPT models (e.g., *GPT-4o*), which are widely known through the ChatGPT web application, and reasoning models (e.g., *o1* or *o3-mini*), which represent a newer model type designed for more structured and logical reasoning[2]. The distinction between these two types of models can be helpfully framed using the dual-system theory of human cognition [57]. According to this theory, human thinking consists of two systems: System 1, which is fast, intuitive, and associative, and System 2, which is slow, analytical, and deliberate [58–61].

*GPT models* are more closely aligned with System 1 thinking [57]. They are highly effective at generating fluent, contextually appropriate responses with remarkable speed, relying heavily on statistical associations learned during training. However, these models tend to struggle with tasks requiring rigorous logical reasoning or multi-step problem solving (e.g., [17,62]). This is due in part to their output generation process, which is strictly forward: each next token is generated based solely on the preceding tokens, without internal planning or revision [63]. As such, GPT models resemble the fast, intuitive, and associative nature of System 1 thinking.

*Reasoning models*, in contrast, are explicitly designed to emulate System 2 thinking [57]. While the exact details of their training and inner workings have not been publicly disclosed, OpenAI has stated that these models were trained using reinforcement learning to perform complex reasoning by efficiently using chain-of-thought—that is, step-by-step reasoning [20,64]. In addition to this explicitly stated optimization for reasoning, findings by McCoy et al. [65] suggest that OpenAI's reasoning models also underwent a substantial amount of training on next-token prediction. However, unlike GPT models, reasoning models are designed to internally generate and process reasoning tokens—intermediate steps that allow them to emulate "thinking" before producing a final answer. These internal reasoning tokens play a key role in helping the model break down a problem and explore different reasoning paths (i.e., chain-of-thoughts) before responding, akin to exploration and exploitation of reasoning paths within tree-of-thought prompting [66]. This process is similar to how a person might silently work through a difficult problem before stating a final solution. This way, reasoning models seem to approximate more closely the slow, analytical, and deliberate nature of System 2 thinking.

---

[2] It is important to recognize that terms such as "understanding" and "reasoning", though commonly used in the literature to describe the behavior of LLMs, are anthropomorphisms and should be interpreted with caution [56]. Apparent instances of understanding, reasoning, or specific cognitive abilities in LLM outputs reflect human interpretations of text produced by computational mechanisms that differ fundamentally from human cognitive processes.

*3. Prompt engineering*

The sensitivity of LLMs to the phrasing of input prompts represents both a limitation and an opportunity, highlighting the importance of prompt engineering as a means of shaping model outputs [67,68]. Effective prompting is essential for optimizing interactions with LLMs, as it enables users to elicit outputs that align with specific informational or stylistic objectives.

For GPT models, prompting was found to have a substantial impact on performance across many tasks. Specifically, OpenAI [69] states that a "GPT model is like a junior co-worker—they will perform best with explicit instructions to create a specific output", highlighting the necessity of clear instruction via prompting to achieve desirable output from an LLM. Prompting techniques that are often mentioned and used in the literature include *chain-of-thought prompting* and *few-shot prompting*.

*Chain-of-thought* (CoT) prompting aims to enhance LLMs' capability to solve multi-step problems by encouraging them to generate a sequence of intermediate reasoning steps prior to arriving at a final answer [70]. Prompting LLMs to think or reason "step-by-step" has been shown to improve the accuracy and quality of their responses across various tasks (e.g., [19,70–72]). Building on this foundation, tree-of-thoughts prompting introduces branching and backtracking between multiple reasoning paths [66], while graph-of-thoughts prompting models reasoning as an arbitrary graph that allows transformations such as merging and refining thoughts [73].

*Few-shot prompting* provides a LLM with multiple example input–output pairs to guide its response on a new task—often leading to performance improvements on certain tasks—with single-shot prompting being a special case that uses only one example [74]. These prompting techniques can also be combined; for instance, few-shot (or single-shot) CoT prompting augments the provided examples by including intermediate reasoning steps within each example. This combined approach has been shown to further enhance LLM performance across various tasks (e.g., [70,72]).

For reasoning models, it is advised to use straightforward prompts rather than prompt engineering techniques such as CoT, which may hinder rather than improve performance [69]. As OpenAI [20] notes, such reasoning models are "like a senior co-worker—you can give them a goal to achieve, and trust them to work out the details." They typically perform well without few-shot examples or explicit step-by-step instructions; examples should only be added when there are more complex requirements for the output.

**C. Physics-related performance of LLMs**

The public release of ChatGPT in November 2022, followed by subsequent releases of further LLMs, has prompted extensive research into their apparent domain-specific conceptual understanding and problem-solving capabilities, in physics as well as in other disciplines. This kind of research is valuable for physics education, particularly, as there seems to be a connection between LLMs' problem-solving and assessment capabilities [13]. Overall, research on LLMs in physics education has demonstrated significant advancements in these models' performance in answering conceptual questions and their apparent problem-solving capabilities over the short period since they gained widespread attention.

*1. Apparent conceptual understanding*

In an early case study, Gregorcic and Pendrill [75] posed a basic conceptual physics question to *ChatGPT-3.5*. Although the model generated linguistically sophisticated responses, they were often inconsistent and unreliable, reflecting a limited capability to reason based on fundamental physics concepts. When dos Santos [76] posed the same conceptual question to *ChatGPT-4*, the model produced a fully correct and detailed explanation.

Subsequent research has evaluated LLMs' apparent conceptual understanding using established concept inventories. West [23], for instance, assessed *ChatGPT-3.5* and *ChatGPT-4* using the Force Concept Inventory (FCI). The results showed that *ChatGPT-3.5* performed comparably to a typical first-semester college physics student (consistent with findings of Kortemeyer [77]), while *ChatGPT-4* demonstrated performance approaching expert levels. Similarly, Tong et al. [14] reported a significant improvement in accuracy from *ChatGPT-3.5* to *ChatGPT-4* on tasks drawn from both the FCI and the Conceptual Survey of Electricity and Magnetism (CSEM), provided that the tasks did not involve visual information.

With the advent of *ChatGPT-4V*, which can process visual input, researchers began exploring its capability to interpret physics-specific visual data. Polverini and Gregorcic [78] evaluated the model's performance on the Test of Understanding Graphs in Kinematics (TUG-K), finding that while *ChatGPT-4V* frequently proposed effective solution strategies and demonstrated sound reasoning, it often failed to accurately extract information from graphs, resulting often in incorrect final answers. Similarly, Aldazharova et al. [79] examined *ChatGPT-4V*'s performance on the FCI, observing strong overall results but also difficulties with items requiring figure interpretation and spatial reasoning. Kortemeyer et al. [80] investigated *GPT-4o*'s multilingual and multimodal conceptual understanding on multiple physics concept inventories covering a wide range of different physics subjects. Their findings indicate unequal performances across subjects and languages, however, *GPT-4o* was found to outperform average post-instruction undergraduate students in almost all subjects. Additionally, performance on purely text-based items exceeded performance on items requiring visual interpretation.

Chapagain et al. [81] tested multiple LLMs on a final higher secondary education physics exam in Nepal that focused on conceptual understanding. *GPT-4o* outperformed all other tested LLMs, achieving 90% of the total score. Beyond understanding of general physics, Holmes et al. [82] explored the capability of several LLMs, including *GPT-3.5* and *GPT-4*, to answer multiple-choice questions on radiation oncology physics—a highly specialized area. Their results showed that *GPT-4* not only outperformed the other tested LLMs but also exceeded the performance of trained medical physicists.

*2. Apparent problem-solving capabilities*

Successfully solving physics problems requires more than conceptual understanding; it also demands knowing how and when to apply conceptual knowledge to solve problems. This difference in complexity is clearly demonstrated in the study by Yeadon et al. [83] which evaluated *ChatGPT-3.5 Turbo* on 1,337 physics exam tasks spanning various educational levels. These tasks included both conceptual questions as well as well-defined physics problems. Their findings revealed that while *ChatGPT-3.5 Turbo* consistently performed well on conceptual questions at earlier educational stages, its performance declined as the content became more advanced, with particularly weak results on the well-defined physics problems. As a result, several studies have moved beyond assessing apparent conceptual understanding, aiming instead to evaluate the apparent problem-solving capabilities of LLMs.

Early investigations into the performance of LLMs on well-defined physics problems yielded mixed results. For example, Liang et al. [84] found that *ChatGPT-3* correctly solved 16 out of 20 simple, well-defined mechanics problems, though occasional computational errors were noted. Similarly, López-Simó and Rezende [85] demonstrated that *ChatGPT-3.5* correctly solved 7 out of 10 simple, well-defined single-step problems but consistently failed on a multi-step problems due to arithmetic errors and misapplications of physics concepts. In another case study, dos Santos [76] compared *ChatGPT-3.5* and *ChatGPT-4* on a single multi-step problem. While *ChatGPT-3.5* failed to grasp the question, *ChatGPT-4* successfully solved it, applying the correct concepts and procedures. However, Kieser and Wulff [86] found that *ChatGPT-4* correctly solved a well-defined multi-step mechanics problem in only 5 out of 10 attempts. Notably, they observed that *ChatGPT-4*'s problem-solving process indicated by the solutions' structures generally aligned with established problem-solving process models—except for the final phase of evaluating the solution, which was consistently omitted.

Going beyond individual case studies, Wang et al. [16] conducted a large benchmark study involving multiple LLMs (e.g., *LLaMA* models, *Claude2*, *GPT-3.5 Turbo*, *GPT-4*, *GPT-4 Turbo*) and various prompting techniques (e.g., no prompting, zero-shot CoT, few-shot CoT) across nearly 300 well-defined college-level physics problems. Proprietary models consistently outperformed open-source ones, with *GPT-4* demonstrating the strongest overall performance. However, it still fell short of human benchmarks, leading the authors to conclude that LLMs' mastery of physics problem solving remains limited. Common errors included miscalculations, flawed causal reasoning, and difficulties in decomposing problems into subproblems. Interestingly, no single prompting technique proved universally effective; rather, different techniques reduced or exacerbated different types of errors. In another benchmark study, Feng et al. [18] tested multiple LLMs (including latest reasoning models such as *o3-mini*, *o1-mini*, and *DeepSeek-R1*) on 1,297 high-level physics problems drawn from physics PhD qualifying exams. Their findings revealed that reasoning models notably outperform general-purpose LLMs, with *o3-mini* outperforming the other models with an overall accuracy of 59.9%. Key errors identified were among others the reliance on incorrect assumptions, difficulties in handling multimodal data, and calculation errors.

While most studies have focused on well-defined problems typical of school and university contexts, some have begun to examine LLM performance on more ill-defined physics problems. For example, Wang et al. [19] assessed *ChatGPT-4*'s performance on 40 engineering physics problems, including both well-defined and ill-defined ones. The model correctly solved 62.5% of the well-defined problems but only 8.3% of the ill-defined ones. Key errors in the latter included not only calculation mistakes but also inaccurate modeling and implausible assumptions. The authors further found that CoT prompting led to a modest increase in accuracy. In another study, Sirnoorkar et al. [15] compared *ChatGPT-3.5* and *ChatGPT-4o* on a single ill-defined physics problem, finding that *ChatGPT-4o* notably outperformed its predecessor in terms of conceptual accuracy. A noteworthy finding was that both models included a variety of detailed assumptions in their solutions—an aspect students often struggle with when tackling ill-defined problems. Overall, these results suggest that ill-defined problems remain especially challenging for LLMs, however, newer models seem to show improvements.

As the aforementioned studies show, recent LLMs—particularly *GPT-4* models—show marked improvement in solving well-defined physics problems. To further assess their capabilities, researchers have begun testing LLMs on more challenging problems, such as those from physics competitions (see [87]). For example, Borovský et al. [88] reported that *GPT-4* and *Claude* successfully solved an advanced problem from the regional stage of the Slovak Physics Olympiad. Likewise, Athiwaratkun [89] tested *GPT-4* on a well-defined problem from the International Physics Olympiad 2011 and reported a score of 4.4 out of 10—a respectable result by Olympiad standards. In a large-scale benchmark study, He et al. [17] evaluated several LLMs (including *GPT-4* and *GPT-4V*) on 8,476 problems from international-level mathematics and physics competitions. *GPT-4V* was the top-performing model on the physics problems but still achieved only 10.7% of the total possible score. As in previous studies, proprietary models outperformed open-source ones (cf. [16]).

In sum, recent research highlights rapid improvements in LLMs' apparent problem-solving capabilities over a relatively short period (e.g., [14,15]). Proprietary models consistently outperform open-source alternatives (e.g., [16–18]), and performance varies notably across LLMs and problem type (i.e., well-defined vs. ill-defined). While newer models demonstrate strong performance on conceptual questions and well-defined physics problems, significant challenges remain for ill-defined problems [15,19] and Olympiad-level problems [17,89]. Research on the effectiveness of prompting techniques remains inconclusive, suggesting minimal to no impact on overall performance [16,19]. Furthermore, little is known about how the newest generation of reasoning models would perform on high-level physics problems, indicating a key area for future investigation.

**D. The present study**

Physics Olympiad problems demand not only deep conceptual understanding but also multi-step reasoning and the precise application of physics concepts through mathematics. This complexity makes them well-suited for evaluating apparent problem-solving capabilities of contemporary LLMs, which have already demonstrated strong performance on conceptual physics questions and simple problems. However, prior research has primarily focused on final answer accuracy, offering limited insight into the problem-solving processes reflected in the complete solutions— particularly the intermediate steps and partial results that contribute to overall problem-solving success. Moreover, the effectiveness of prompting techniques remains inconclusive, and little is known about the performance of the latest reasoning-optimized models in this context. Against this backdrop, the present study aims to systematically compare the performance of a general-purpose LLM under multiple prompting techniques and a reasoning-optimized LLM, based on their respective solutions to advanced Physics Olympiad problems. We specifically ask the following research question:

> To what extent does the performance of a general-purpose LLM (*GPT-4o*; under varying prompting techniques), a reasoning-optimized LLM (*o1-preview*), and actual Physics Olympiad participants compare in solving advanced problems from the Physics Olympiad?

# III. METHODS

## A. Selection of Physics Olympiad problems

We systematically analyzed all 105 physics problems used at any stage of the most recent German Physics Olympiads (2022, 2023, and 2024)[3] in terms of both their subject matter (e.g., mechanics, optics) and their mode of presentation (e.g., text only, text with required illustrations or tables). Given that mechanics problems were the most prevalent and that prior research has shown that GPT models struggle to extract information from visualizations (e.g., [78,80]), we chose to focus on mechanics problems that were presented almost entirely in text form. As a result, we selected six text-based mechanics problems from the Olympiad with minimal to no reliance on visual information. Table I summarizes some details about these problems, including the involved physics concepts. It is important to note that higher-stage problems are not necessarily more difficult than lower-stage ones, as each stage includes simpler as well as more challenging problems. Furthermore, the full set of problems, translated from German to English by the authors, is available in the Supplemental Material (Part A; see Ref. [111]). For the purposes of this study, two problems (ICE and ROC) were slightly modified to remove illustrations that contained essential information for solving the problems[4].

TABLE I. Overview of the selected Physics Olympiad problems utilized in this study, including the competition stage and year they were employed, the key physics concepts involved, and the maximum achievable points. It also includes a difficulty ranking from 1 (simplest) to 3 (most difficult), based on the authors' estimates of the required abilities as well as the number and complexity of steps needed to solve each problem. An asterisk next to the problem name indicates that the problem was slightly modified to circumvent the reliance on visual representations. It is important to note that ICE and ROC are progressive problems in that they consist of multiple interconnected subproblems built around a common scenario, with certain subproblems depending on answers or information from earlier subproblems.

| Problem name | Stage, year | Involved physics concepts | Points | Difficulty |
|---|---|---|---|---|
| [HEL] Helicopter on Mars | 1st stage, 2024 | momentum, induced flow, force equilibrium, gravitational acceleration on a planet | 10 | 2.0 |
| [ICE] Capsizing Iceberg | 1st stage, 2022 | Archimedes' principle, change in potential energy | 10 | 2.0 |
| [EXO] Fall on Exoplanet* | 2nd stage, 2022 | free fall, gravitational acceleration on a planet | 5 | 1.0 |
| [ROC] Rocket Launches and Satellites* | 2nd stage, 2023 | elastic collision, momentum, Newton's second law, gravitational force, centripetal force, conservation of mechanical energy, Kepler's third law | 14 | 3.0 |
| [INS] Insect Hunting | 3rd stage, 2023 | classical Doppler effect with moving receiver and source | 5 | 2.5 |
| [SLE] Sled Pulling | 4th stage, 2024 | sliding friction, component decomposition of forces, force equilibrium | 4 | 1.5 |

---

[3] To clarify, the term German Physics Olympiad 2022 refers to the annual national selection process, which comprises four consecutive stages—collectively known as the Physics Olympiad—starting in April 2021 and culminating in the selection of participants for the International Physics Olympiad 2022 in spring 2022.

[4] Specifically, subtask (c) of the ICE problem required estimating the width of an iceberg from an image. For this study, we rephrased the subtask to provide this information explicitly while also ensuring that no LLM-generated solution was awarded points for extracting the information from the image. Similarly, subtask (b) of the ROC problem involved extracting data from two graphs; for the purposes of this study, we omitted this subtask entirely.

**B. Generation of solutions using different prompting techniques**

We employed two advanced LLMs developed by OpenAI. More precisely, we used the general-purpose GPT model *gpt-4o-2024-08-06* (in the following just referred to as *GPT-4o*; accessed via the OpenAI API with a temperature setting of 0.7) and the reasoning model *o1-preview* (accessed via the ChatGPT web interface). At the time this study commenced, the *o1-preview* model had just been released as OpenAI's first reasoning model and was available exclusively to ChatGPT Plus users.

For *GPT-4o*, we implemented four prompting techniques: *no prompting*, *general prompting*, *Chain-of-Thought (CoT) prompting*, and *single-shot CoT prompting*. The structure of the corresponding prompts follows a modular design, where each prompt consists of one or more subprompts as well as the problem text (see FIG. 1), all provided to the LLMs in German language:

- **No prompting:** Only the raw *problem text* of each Olympiad problem was provided.
- **General prompting:** A *general prompt* was placed before the problem text, mirroring the guidance typically provided to students alongside Olympiad problems (see FIG. 2 for the exact wording).
- **CoT prompting:** In addition to the *general prompt*, a *CoT prompt* was introduced (see FIG. 2 for the exact wording), instructing the LLM to follow a step-wise reasoning approach aligned with established physics problem-solving process models. These models typically consist of four sequential phases (problem representation, strategy selection, strategy execution, and evaluation) which were considered in the prompt.
- **Single-shot CoT prompting:** In addition to the *general prompt* and the *CoT prompt*, the LLM was provided a *single-shot prompt*, describing a worked example illustrating a structured problem-solving process (see FIG. 2 for the exact wording).

For *o1-preview*, only the *general prompting* technique was tested as OpenAI [69] specifically stated that their reasoning models perform best with straightforward prompts.

To account for the inherent randomness in LLMs' outputs, we generated 20 independent solutions for each of the six physics problems across all five considered prompting configurations. This included four different prompting techniques for *GPT-4o* and one for *o1-preview*. Independent solutions were obtained either through separate API calls for *GPT-4o* or by opening a new chat window for each solution attempt in the *o1-preview* scenario. In total, this resulted in $6 \times 20 \times 5 = 600$ solutions generated by the LLMs that were saved for further processing.

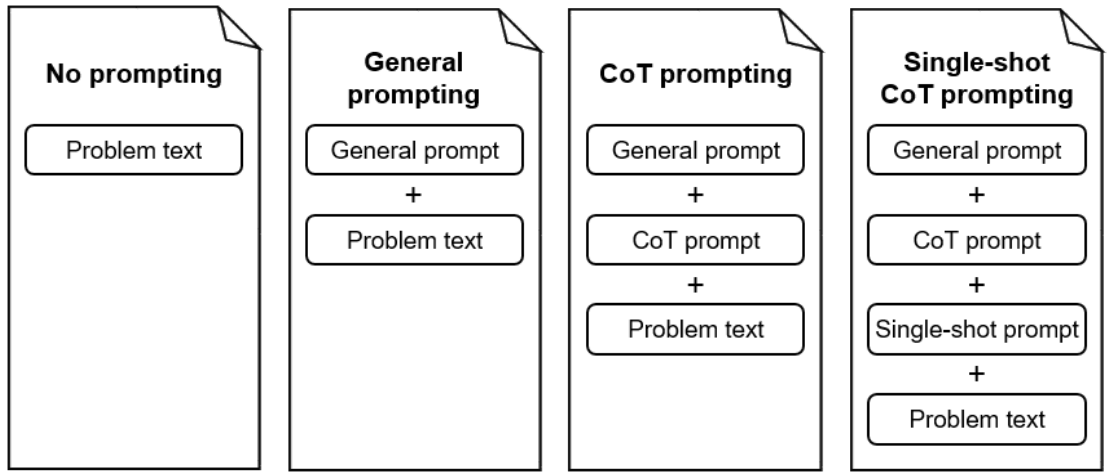

**FIG. 1.** Illustration of the modular design of the utilized prompts corresponding to the different prompting techniques employed. From left to right, the prompts become increasingly complex as a new subprompt is gradually added. The exact wording of the subprompts is given in FIG. 2.

> **General prompt:**
>
> "Solve the following physics problem. Provide a detailed explanation of your solution so that each step can be followed. Explicitly state any approximations and assumptions you make, even if they are not explicitly requested in the problem statement."
>
> **CoT prompt:**
>
> "Start by stating the assumptions and idealizations relevant to the problem. Next, identify potentially relevant physics concepts. Use these concepts to solve the problem, and finally, assess the plausibility of your solution."
>
> **Single-shot prompt:**
>
> "Here is a sample problem with a solution illustrating this approach:
>
> A small mass moves along a path that includes a vertical loop. The mass starts from a height above the highest point of the loop. Determine the minimum starting height required for the mass to complete the loop without falling down.
>
> First, assumptions and idealizations must be made. The small mass is treated as a point mass, which means that the motion is frictionless and rotation does not need to be considered. Furthermore, the loop is assumed to be a circular path.
>
> Next, relevant physics concepts must be identified. First, the law of energy conservation holds, meaning that the total energy of the mass, consisting of potential and kinetic energy, remains constant. Additionally, a centripetal force must act in the loop since the mass follows a circular trajectory. In particular, at the highest point of the loop, the centripetal force must be exactly equal to the gravitational force acting on the mass.
>
> The problem should then be solved using the identified physics concepts. Initially, the mass is at rest at height $h$ and therefore has only potential energy $E\_pot,1 = mgh$. At the highest point of the loop, the mass has both potential and kinetic energy, given by $E\_pot,2 + E\_kin = mg*2R + 1/2mv^2$, where $2R$ is the diameter of the loop and $v$ is the velocity of the mass at the highest point of the loop. Applying the law of energy conservation, we obtain $E\_pot,1 = E\_pot,2 + E\_kin$, so $mgh = mg*2R + 1/2mv^2$. Dividing both sides by $mg$ yields $h = 2R + v^2/(2g)$. The velocity $v$, or rather $v^2$, can now be determined using the fact that at the highest point of the loop, the gravitational force $F\_G = mg$ is equal to the centripetal force $F\_Z = mv^2/R$, leading to $mg = mv^2/R$. Rearranging for $v^2$ gives $v^2 = gR$. Substituting this into the equation for $h$ yields $h = 2R + v^2/(2g) = 2R + gR/(2g) = 2R + R/2 = 5R/2$. Thus, the minimum starting height is $2.5R$.
>
> Finally, the result should be checked for plausibility, for example, through a unit analysis or logical reasoning. The solution $2.5R$ has the dimension of a length, just like the required height. Moreover, the required height $2.5R$ is greater than the loop height $2R$, which is also reasonable based on energy conservation. Therefore, the result appears to be plausible.
>
> Now, solve the following problem: "

**FIG. 2.** Wordings of the general prompt, CoT prompt and single-shot prompt (translated from German to English by the authors).

## C. Scoring of solutions

The scoring of the LLM-generated solutions was conducted by two raters (a physics graduate student and an experienced physics teacher; both among the authors of this manuscript) based on the official problem-specific scoring schemes used in the Physic Olympiad[5]. These scoring schemes involve criteria not only addressing the correctness of the final answer but also further aspects of the problem-solving process, including the construction of an adequate problem representation, the identification of relevant physics concepts, and their application typically involving mathematics. The concrete scoring schemes are available in the Supplemental Material (Part A; see Ref. [111]). Each generated solution was evaluated by assigning subscores according to these criteria, and subscores per problem were ultimately aggregated to a single score. The primary rater (a graduate physics student) scored 100% of the generated solutions (i.e., a total of 600 solutions), while a second rater (an experienced physics teacher) independently scored a subset of 10% (60 solutions, specifically 10 solutions for each of the six physics problems).

Interrater agreement was evaluated based on the 10% subset of generated solutions that had been independently scored by both raters. Agreement was assessed separately for each problem using the mean absolute difference (MAD) of the relative scores assigned by the two raters to each solution. These values represent the average difference in ratings between the two raters relative to the maximum possible score per problem. The obtained MAD values were as follows: $MAD_{HEL} = 15.2\%$, $MAD_{ICE} = 6.2\%$, $MAD_{EXO} = 6.0\%$, $MAD_{ROC} = 12.1\%$, $MAD_{INS} = 5.0\%$ and $MAD_{SLE} = 4.5\%$, indicating some discrepancies that were generally not substantial. To ensure consistency in scoring, a consensus coding process was carried out on this subset: Any notable differences in scorings, particularly those related to the detailed scoring scheme, were discussed between the two raters until a consensus was reached. Based on the insights gained during the consensus coding process, the primary rater subsequently revisited the remaining 90% of the solutions and revised his ratings to ensure alignment with the consensus-based interpretations of the scoring criteria. The resulting scores were then used for all subsequent analyses.

## D. Analysis of LLMs' apparent problem-solving capabilities

To compare the performance of *GPT-4o* (depending on different prompting techniques) and *o1-preview*, we first aggregated scores at the level of complete six-problem sets to simulate an exam-like setting. Since each of the six problems was solved independently by the LLMs, we constructed artificial "exam responses" by randomly combining one independently generated solution per problem into a single six-problem set. For each of these simulated sets, we computed a total exam score by summing the individual problem scores.

Given that there are $20^6 = 64,000,000$ possible combinations per prompting technique (based on 20 generated solutions for each of the six problems), we sampled $N = 100,000$ such combinations uniformly at random. This sampling procedure approximates the distribution of total exam scores that would result from the LLM solving all six problems together under the given prompting condition multiple times. This enables a robust comparison of average exam-level performance across LLMs and prompting techniques, while remaining computationally tractable. The same procedure was applied to the scored solutions of Physics Olympiad participants[6], as the problems originated from different years and stages of the Physics Olympiad, resulting in varying underlying student populations. Resulting distributions of simulated aggregated scores were inspected visually and differences between distributions were quantified using pairwise independent two-sided Welch's *t*-tests[7] [91] and Cohen's *d* as a measure of effect size [92].

---

[5] Scoring of students' solutions was performed by teachers or experienced jury members using the same scoring schemes, typically in a two-stage process.

Moreover, LLMs' and Physics Olympiad participants' performance was evaluated problem-wise via visual analysis using boxplots. To quantitatively assess differences, pairwise comparisons using two-sided Mann-Whitney *U* tests [93] were also conducted. These tests evaluated whether the problem-wise score distributions suggested significant differences between *GPT-4o* (depending on different prompting techniques), *o1-preview*, and actual Physics Olympiad participants. Based on these quantitative findings, we identified typical errors and fallacies that occurred frequently in LLM-generated solutions.

## IV. RESULTS

### A. Quantitative analysis

Distributions of aggregated scores for *GPT-4o* (across different prompting techniques), the *o1-preview* reasoning model, and Physics Olympiad participants are depicted in FIG. 3. As can be readily observed, Physics Olympiad participants on average perform worse than the LLMs, regardless of the prompting technique used. Among all models, *o1-preview* achieves the highest performance, followed by the different prompting techniques based on *GPT-4o*. In addition to these differences in the average performance, notable differences in score variance are also evident: while student performance exhibits a substantially higher variance, the variance of scores produced by *GPT-4o* is considerably smaller, and the scores of *o1-preview* are even more narrowly distributed.

---

[6] The number of student solutions per Olympiad problem varies notably depending on the competition stage in which the problem was used, as the number of participants decreases in later stages. Moreover, scores for individual problem solutions from the second stage onward should be interpreted with caution, as these problems were always part of a timed multi-problem examination. Consequently, a student's score on a single problem may not fully reflect their problem-solving ability, as they may have skipped the problem or prioritized others.

[7] As shown later in FIG. 3, the resulting distributions are approximately normal but exhibit partially unequal variances. Therefore, Welch's *t*-test is recommended [90].

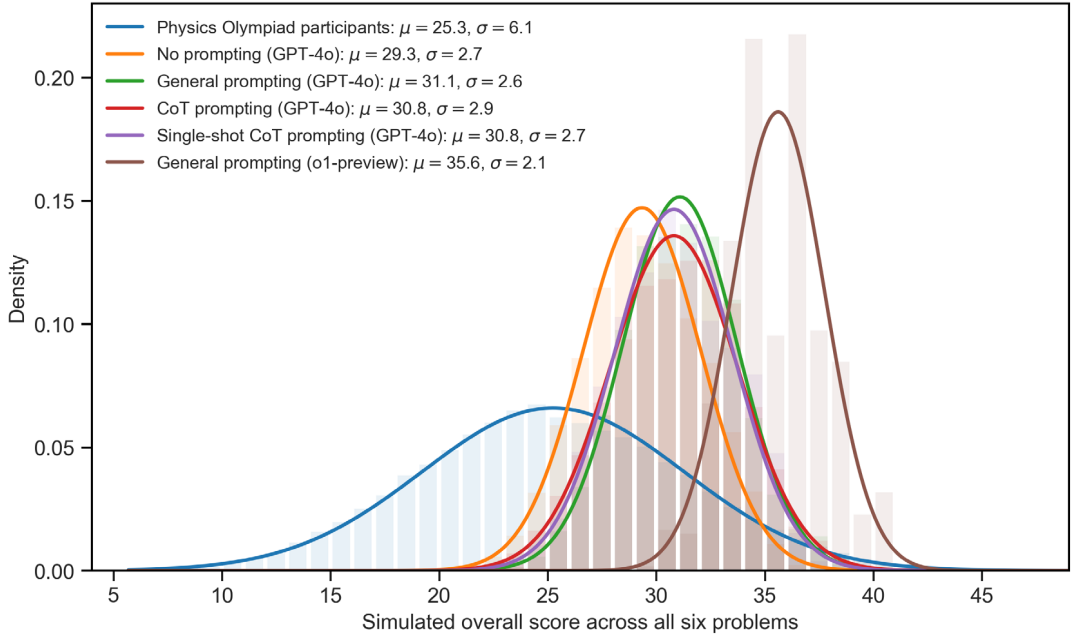

**FIG. 3.** Distributions of simulated aggregated scores for Physics Olympiad participants, *GPT-4o* (involving different prompting techniques) and *o1-preview* (general prompting). To simulate a full-exam setting, scores across all six problems were aggregated. Since problem solutions were generated independently for each problem, exam-like performance was approximated by sampling from the Cartesian product of the scored responses—that is, by randomly selecting combinations of one response per problem from the available solutions. This approach reflects how a LLM might have performed if required to solve all six problems in a single attempt. As the resulting score distributions were approximately normal, fitted normal curves (characterized by their mean $\mu$ and standard deviation $\sigma$) were overlaid to facilitate visual comparisons. Note that a maximum of 48 points was achievable across the six problems.

To statistically assess the observed performance differences, pairwise Welch's *t*-tests were conducted. Owing to the high number of simulated exam samples ($N = 100{,}000$), nearly all comparisons between LLM setups yielded statistically significant results. An exception (no significant difference) was found only in the comparison between CoT prompting and single-shot CoT prompting using *GPT-4o*.

To complement these findings and provide an interpretable and practical measure of the magnitude of mean differences, we also examined effect sizes using Cohen's *d*. The results, as summarized in TABLE II and interpreted according to Sawilowsky [94] reveal that the *o1-preview* reasoning model in combination with general prompting substantially outperforms all other tested LLM setups and the Physics Olympiad participants, with very large to huge effect sizes. Moreover, Physics Olympiad participants consistently performed worse than all tested *GPT-4o* setups, irrespective of the employed prompting technique, with large effect sizes. In contrast, comparisons among the *GPT-4o* prompting setups (i.e., general prompting, CoT prompting, and single-shot CoT prompting) show no or only small effect sizes, indicating comparable problem-solving performance. Additionally, using *GPT-4o* with no prompting tended to result in slightly lower scores than the other prompting techniques, reflected in medium effect sizes.

In a second step, problem-solving performance was evaluated on a problem-by-problem basis through visual inspection using boxplots (see FIG. 4). The comparison included Physics Olympiad participants, *GPT-4o* (across various prompting techniques), and *o1-preview* (with general prompting). A notable observation is the substantially higher variance in scores among Physics Olympiad participants, as indicated by wider interquartile ranges and whiskers, compared to the relatively narrow distributions observed for the LLM-generated solutions. Furthermore, the score distributions across different prompting techniques using *GPT-4o* appear largely similar, suggesting that prompt variation had limited impact on overall performance. In contrast, the score distribution for *o1-preview* generally differs qualitatively from those of *GPT-4o*, although it does not consistently outperform them. To statistically quantify these differences, pairwise comparisons between all groups were conducted using Mann–Whitney $U$ tests. The resulting *p*-values are reported in TABLE III.

**TABLE II**. Comparison of the overall score distributions depicted in FIG. 3 using Cohen's *d* as a measure of (absolute) effect size.

|  | no prompting (*GPT-4o*) | general prompting (*GPT-4o*) | CoT prompting (*GPT-4o*) | single-shot CoT prompting (*GPT-4o*) | general prompting (*o1-preview*) |
|---|---|---|---|---|---|
| Physics Olympiad participants | 0.87 | 1.25 | 1.17 | 1.19 | 2.28 |
| no prompting (*GPT-4o*) |  | 0.65 | 0.52 | 0.54 | 2.57 |
| general prompting (*GPT-4o*) |  |  | 0.10 | 0.10 | 1.88 |
| CoT prompting (*GPT-4o*) |  |  |  | 0.00 | 1.87 |
| single-shot CoT prompting (*GPT-4o*) |  |  |  |  | 1.96 |

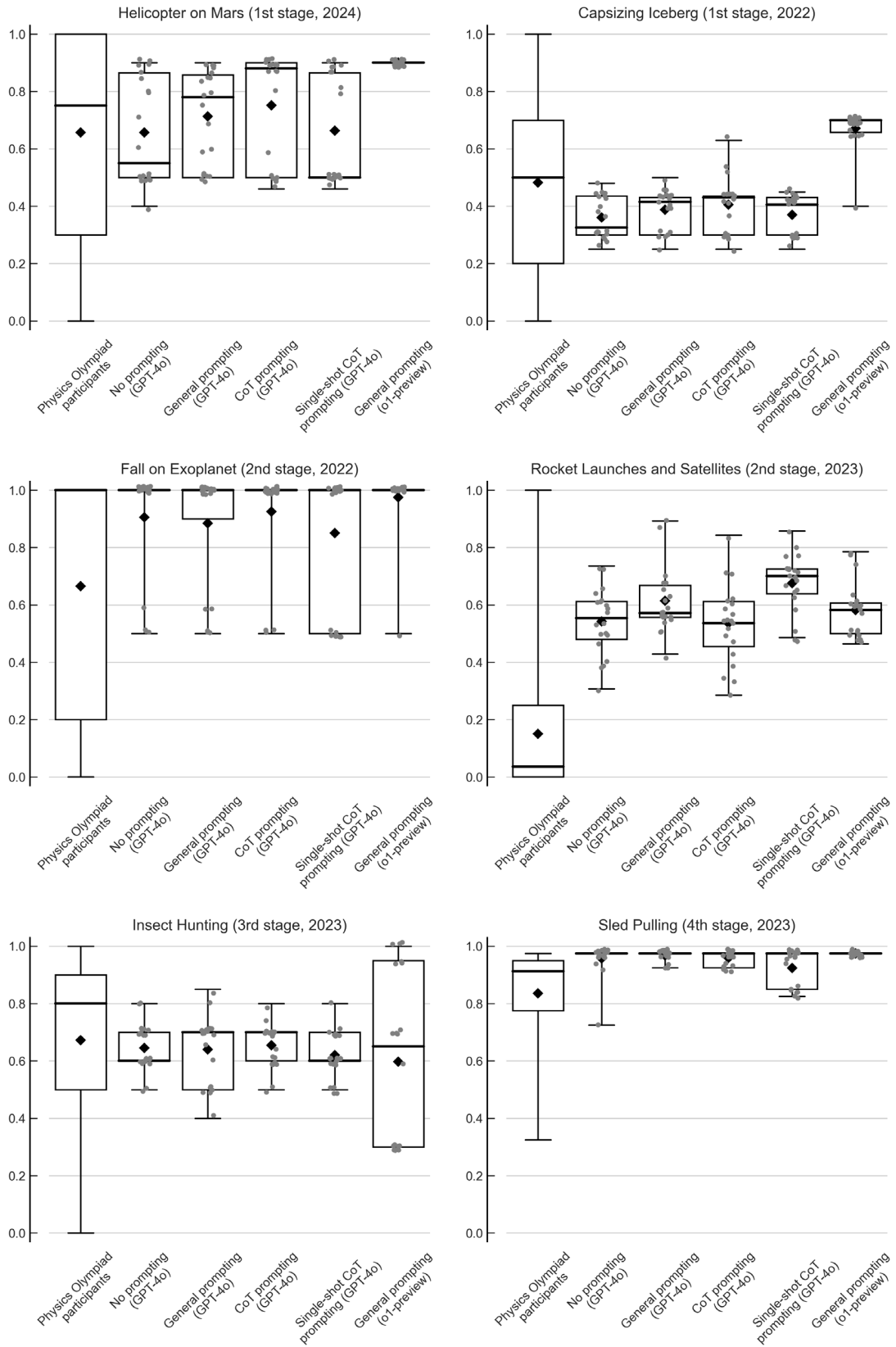

**FIG. 4.** Boxplot visualization of the achieved scores across the examined Physics Olympiad problems. For each problem, six boxplots represent the score distributions of solutions provided by Physics Olympiad participants and those generated by LLMs. The diamond-shaped markers indicate the average scores, while the whiskers extend to encompass the full range of achieved scores.

**TABLE III.** Comparison of the scores achieved by Physics Olympiad participants, *GPT-4o* (using different prompting techniques), and *o1-preview* (using general prompting) for each of the six Olympiad problems. The reported values represent *p*-values obtained from two-sided Mann-Whitney *U* tests for pairwise comparisons, with values in bold indicating statistical significance (i.e., $p < 0.05$).

| | no prompting (*GPT-4o*) | general prompting (*GPT-4o*) | CoT prompting (*GPT-4o*) | single-shot CoT prompting (*GPT-4o*) | general prompting (*o1-preview*) | no prompting (*GPT-4o*) | general prompting (*GPT-4o*) | CoT prompting (*GPT-4o*) | single-shot CoT prompting (*GPT-4o*) | general prompting (*o1-preview*) |
|---|---|---|---|---|---|---|---|---|---|---|
| | \multicolumn{5}{c}{Helicopter on Mars [HEL]} | \multicolumn{5}{c}{Capsizing Iceberg [ICE]} |
| Physics Olympiad participants | 0.48 | 0.88 | 0.73 | 0.54 | **0.03** | 0.13 | 0.30 | 0.47 | 0.20 | **<0.01** |
| no prompting (*GPT-4o*) | | 0.40 | 0.13 | 0.98 | **<0.01** | | 0.35 | 0.21 | 0.81 | **<0.01** |
| general prompting (*GPT-4o*) | | | 0.26 | 0.44 | **<0.01** | | | 0.67 | 0.54 | **<0.01** |
| CoT prompting (*GPT-4o*) | | | | 0.12 | **<0.01** | | | | 0.33 | **<0.01** |
| single-shot CoT prompting (*GPT-4o*) | | | | | **<0.01** | | | | | **<0.01** |
| | \multicolumn{5}{c}{Fall on Exoplanet [EXO]} | \multicolumn{5}{c}{Rocket Launches and Satellites [ROC]} |
| Physics Olympiad participants | **0.01** | **0.02** | **<0.01** | **0.04** | **<0.01** | **<0.01** | **<0.01** | **<0.01** | **<0.01** | **<0.01** |
| no prompting (*GPT-4o*) | | 0.77 | 0.74 | 0.42 | 0.17 | | 0.14 | 0.85 | **<0.01** | 0.42 |
| general prompting (*GPT-4o*) | | | 0.52 | 0.59 | 0.09 | | | 0.07 | **0.03** | 0.42 |
| CoT prompting (*GPT-4o*) | | | | 0.27 | 0.31 | | | | **<0.01** | 0.35 |
| single-shot CoT prompting (*GPT-4o*) | | | | | **0.04** | | | | | **0.01** |
| | \multicolumn{5}{c}{Insect Hunting [INS]} | \multicolumn{5}{c}{Sled Pulling [SLE]} |
| Physics Olympiad participants | 0.09 | 0.10 | 0.10 | **0.04** | 0.37 | **<0.01** | **<0.01** | **<0.01** | **0.02** | **<0.01** |
| no prompting (*GPT-4o*) | | 0.86 | 0.58 | 0.37 | 0.68 | | 0.97 | 0.21 | 0.05 | 0.08 |
| general prompting (*GPT-4o*) | | | 0.85 | 0.39 | 0.56 | | | 0.17 | **0.03** | 0.08 |
| CoT prompting (*GPT-4o*) | | | | 0.15 | 0.57 | | | | 0.24 | **<0.01** |
| single-shot CoT prompting (*GPT-4o*) | | | | | 0.85 | | | | | **<0.01** |

## B. Detailed analysis of LLM errors

In the *Helicopter on Mars* problem, only the *o1-preview* model performed significantly differently from all other LLMs and from the Physics Olympiad participants (see TABLE III), achieving a remarkably consistent score of 9 out of 10 across all 20 generated solutions. Notably, none of the LLM-generated responses received full credit, whereas some student solutions did. The point most frequently deducted concerned the failure to "propose a relationship for the mass of gas pushed down per unit of time by the rotor"—a theoretically necessary step for deriving that the upward force is proportional to both the atmospheric density and the square of the rotor's frequency. Rather than deriving this relationship, the LLM-generated responses tended to assume it without justification. A further common error, observed in *GPT-4o*-generated solutions but not in those generated by *o1-preview*, was the omission of the relevant gravitational accelerations for Earth and Mars in the final derived equation. Notably, although the problem statement explicitly instructed to first determine the gravitational acceleration on Mars—and most *GPT-4o*-generated solutions did compute this value correctly (albeit occasionally confusing Mars' radius with its diameter)—this value was typically not incorporated into the subsequent reasoning. In contrast, a careful student might reasonably question the purpose of such a calculation if its result is not utilized in the solution process. The omission of the gravitational accelerations in the final answer appeared to stem from two main types of reasoning errors: either from directly equating the upward forces on Earth and Mars, or from correctly noting that the upward force on each planet must exceed the helicopter's weight, but then mistakenly assuming that the weights are identical on both planets. This latter assumption is particularly questionable given that most solutions correctly calculated the gravitational acceleration on Mars, which clearly differs from that on Earth, implying that the weights must also differ.

In the *Capsizing Iceberg* problem, *o1-preview* again outperformed *GPT-4o* (independent of prompting) and the Physics Olympiad participants (see TABLE III), with virtually no variability across its generated solutions (see FIG. 4). In more detail, all LLM-generated solutions correctly answered the first subtask—determining the proportion of the iceberg submerged below the water surface. However, in the next subtask, none of the LLM-generated solutions considered the potential energy of the displaced water when evaluating the tipping behavior of the iceberg. Although the change in potential energy of the iceberg itself was consistently addressed, *GPT-4o* solutions frequently failed to account for the fact that a portion of the iceberg lies below the water surface—despite having correctly established this in the first subtask—thus misrepresenting the same geometry in their energy analysis. In contrast, *o1-preview* nearly always incorporated this aspect correctly. A further key difference emerged in the derivation of the condition for maximum energy release due to capsizing. Theoretically, it is sufficient to accurately model either the energy change of the iceberg or of the displaced water to obtain the correct maximization condition via a simple consideration of the derivative, as leaving one out effects only a constant factor. While this would have allowed *GPT-4o* to reach a correct answer despite certain inaccuracies in prior intermediate results, it generally failed to do so, whereas *o1-preview* reliably produced the correct condition. Finally, in the concluding subtask, *GPT-4o* frequently stopped after substituting known values into the relevant equations without computing the final numerical result. This pattern did not occur in *o1-preview*'s outputs, which consistently followed through to a complete numerical result.

In the *Fall on Exoplanet* problem, both *GPT-4o* and *o1-preview* significantly outperformed actual Physics Olympiad participants (see TABLE III). A small ceiling effect was observed, with 81 out of 100 LLM-generated solutions receiving the maximum possible score (see FIG. 4), suggesting that the LLMs encountered little difficulty with the problem. Errors occurred only in two specific forms: in three instances, the correct necessary equation was derived, but a basic integer calculation failed; in sixteen cases, the LLMs failed to correctly manipulate the relevant equation.

For the *Rocket Launches and Satellites* problem, *GPT-4o* and *o1-preview* again significantly outperformed Physics Olympiad participants (see TABLE III). Students' weaker performance may partly be attributed to the position of the specific problem near the end of the original exam, where time constraints likely influenced the quality of their solutions. Among the tested LLMs and prompting techniques, single-shot CoT prompting using *GPT-4o* yielded the highest scores. This advantage can be traced to the design of the single-shot subprompt, which included a worked example explicitly invoking the key physics principles—centripetal force via gravity and energy conservation—necessary to derive the first and second cosmic velocities, two central parts of the problem. However, none of the LLMs succeeded in correctly solving the first subtask, which involved modeling the deflection of air molecules on a cone-shaped tip. This failure appears rooted in difficulties forming an adequate spatial representation of the physical setup. Interestingly, in the final subtask, *GPT-4o*-generated solutions often followed the given hint that Kepler's laws might be useful and proceeded to apply Kepler's third law to solve the problem. However, the reasoning provided was generally not very clear or detailed. In contrast, *o1-preview*-generated solutions often simply stated the correct final formula without explaining where it came from. In some cases, the responses outlined that the equation could be obtained by integrating the differential equation of radial motion under gravity, derived from the conservation of mechanical energy. However, they consistently skipped the nontrivial mathematical steps required to carry out this integration.

In the *Insect Hunting* problem, there were no major differences in performance between *GPT-4o* and *o1-preview* (see TABLE III and FIG. 4). However, *o1-preview*-generated solutions exhibited greater variability: while they included both the best- and worst-performing LLM-generated solutions, the average performance was comparable to *GPT-4o*. Interestingly, some of the lowest-scoring solutions had reached the correct final result by applying a precise approximation formula for a double application of the Doppler effect under low-velocity conditions. These responses received low scores because they stated the correct result based on an equation that was never derived, leading to minimal credit for the derivation. However, if additionally prompted to explain where this approximation comes from, *o1-preview* provided a clear and correct derivation which would have received full credit. Notably, beside these *o1-preview*-generated solutions using the approximation formula, only other *o1-preview*-generated solutions managed to correctly apply the full double Doppler formula, whereas all *GPT-4o*-generated responses failed at some point during the required manipulations of equations.

For the *Sled Pulling* problem, LLM-generated solutions consistently outperformed those of Physics Olympiad participants (see TABLE III). A pronounced ceiling effect can be observed, with most LLM responses achieving nearly full credits (see FIG. 4). The minor deductions that occurred were due to the failure of all LLM-generated solutions to verify that an identified extremum was indeed a maximum, for example through a second-order-derivative test. Across all tested configurations, there were no meaningful differences in the quality of solutions between LLMs or prompting techniques, suggesting that this particular problem was straightforward for LLMs regardless of how they were prompted.

## V. DISCUSSION

### A. Apparent physics problem-solving capabilities of LLMs

The findings of this study indicate that the tested LLMs (*GPT-4o* and *o1-preview*) demonstrate advanced aaparent problem-solving capabilities on Olympiad-type physics problems, on average exceeding the performance of actual Physics Olympiad participants. Notably, the reasoning-optimized model *o1-preview* almost consistently outperformed both the general-purpose model *GPT-4o* (regardless of the utilized prompting technique) and human participants. In line with prior research (e.g., [16,19]), we observed that the employed prompting techniques had generally little to no impact on *GPT-4o*'s performance. This suggests that improvements in performance may depend more on architectural or training-level enhancements of LLMs rather than on external prompting techniques. While human participants underperformed relative to LLMs on average, their solutions exhibited a wider variability in score, which is also consistent with findings of Kieser et al. [95]. Human responses spanned the full scoring range—from very low to very high— whereas LLM-generated solutions typically fell within a narrower band, suggesting a more consistent performance profile.

Qualitative analysis revealed several recurring error patterns in the LLMs' solutions. Both *GPT-4o* and *o1-preview* frequently asserted nontrivial formulas without offering derivations or justifications in the solutions to three of the six examined problems. This suggests a broader difficulty of LLMs in distinguishing between commonly accepted knowledge (including formulas) and more advanced or problem-specific knowledge typically requiring justification. This behavior is likely a result of the models retrieving memorized formulas from training when encountering familiar contexts ("LLMs as stochastic parrots"; see Ref. [96]), without considering whether such information should be treated as known or should be derived.

Furthermore, in two problems, we also observed that *GPT-4o* demonstrated a lack of integrative reasoning compared to the reasoning model *o1-preview*. *GPT-4o* consistently treated subtasks in isolation, failing to propagate intermediate results, and thereby engaged in local reasoning without maintaining global coherence across the entire solution. Interestingly, this pattern of local reasoning—characterized by isolated subtask processing and a failure to integrate intermediate results—bears resemblance to novice-like problem solving, which has been associated with limited working memory resources (e.g., [97]) and more unstructured problem solving (e.g., [39,98]).

We also observed challenges in symbolic manipulation and numerical computation, particularly in the case of *GPT-4o*, consistent with findings from previous studies (e.g., [16,18,19]). This behavior is understandable, given that LLMs are designed for natural language processing rather than for executing formal symbolic reasoning or numerical computation. It is, in fact, remarkable that LLMs are capable of handling mathematical expressions and quantitative reasoning to any degree, especially in the absence of explicit access to external tools or resources (such as Python or Wolfram Alpha). In our setup, we interacted with *GPT-4o* via the API, without enabling external function-calling tools such as Python or Wolfram Alpha. In contrast, *o1-preview* was accessed via the ChatGPT interface and may have had latent access to external tools. Tool usage is typically indicated explicitly (e.g., through generated code when Python is used), and no such indications were present in the outputs. However, as tool access cannot be disabled in this interface, we cannot entirely rule out the possibility. In one problem, *GPT-4o* frequently omitted the final numerical computation in the last subtask, even after correctly identifying the relevant formula and substituting values. This pattern reflects a form of "lazy" behavior, a term used by Zhao et al. to describe similar tendencies in LLMs [99].

Finally, our results point to persistent weaknesses in modeling physical systems and situations, particularly in contexts that require system thinking and spatial understanding. For example, both *GPT-4o* and *o1-preview* struggled to represent the dynamics of an iceberg capsizing in water and the deflection of air molecules on a conical surface. These difficulties align with known limitations in extracting and integrating visual information [78–80] and may reflect a broader challenge: unlike humans, LLMs lack the capacity to construct and interrogate coherent (mental) models of physical systems.

**B. Limitations and future directions**

This study offers several insights into the apparent physics problem-solving capabilities of LLMs, but it also comes with limitations that in some instances also open avenues for further investigation.

One key limitation concerns the stability of model outputs over time. To strengthen the robustness of our findings, we decided to double the number of generated solutions per problem and prompting technique, generating additional data six weeks after the initial generation using the same prompts and problem formulations. However, for some problems, we observed significantly different output from *GPT-4o* based on rated scores, despite these controlled conditions (see Supplemental Material Part B in Ref. [111]). This discrepancy likely reflects undocumented backend changes to the LLMs, highlighting a broader challenge when working with proprietary, black-box systems. More generally, this relates to the phenomenon of temporal variability in LLMs, where model behavior subtly but meaningfully shifts over time [100]. These changes complicate reproducibility and make it difficult to isolate model behavior from platform changes. Future research should systematically investigate the scope and scale of such temporal effects, both over extended periods and potentially even within shorter timescales (e.g., fluctuations in performance over the course of a single day).

Relatedly, while our study focused on *GPT-4o* and *o1-preview*, many of the observed patterns may or may not generalize to other LLMs. Systematic comparisons with open-source models (such as *LLaMA*) or other proprietary models (such as *Claude* and *Gemini*) could help determine whether our findings are specific to the investigated models or reflect broader characteristics of current LLMs. In this context, the language of the problems also warrants consideration: all problems and generated solutions in this study were in German, which may limit generalizability. A recent study found that *GPT-4o* performs similarly across English and several European languages on physics concept inventories, but showed lower performance for non-Western languages [80]. We therefore hypothesize that cross-linguistic generalization of our findings is plausible within the European language family, but that diminished problem-solving performance of LLMs may be expected for non-Western languages. However, these are just speculations, and there is a need for replication in diverse linguistic contexts to more rigorously test the robustness and generalizability of our findings beyond German.

Another important limitation concerns the potential for task contamination. Prior work shows that LLMs often perform better on datasets that were available during their training period [101]. Given that both LLMs tested in this study were trained on data up to October 2023—and that most problems including solutions (except INS and SLE; see TABLE I) were publicly accessible prior to that—it remains unclear whether the models indeed "engaged in problem solving" or solely reproduced previously seen material. This uncertainty is an additional reason why one needs to refer to the models' behavior as *apparent* problem solving. However, this potential task contamination does not substantially affect the implications of our findings for educational assessment.

Finally, our results offer implications for the use of LLMs in data augmentation—particularly in generating synthetic student solutions for training or evaluation purposes regarding machine learning-based applications or research [95,102]. In our case, the synthetic data (i.e., the LLM-generated solutions) were on average better than actual student solutions and exhibited much lower variability. To better mimic authentic student distributions, future work could explore techniques that lower the average response quality (e.g., through prompting) and increase variability (e.g., through increasing the LLMs' temperature).

## VI. IMPLICATIONS FOR ASSESSMENT

The rapid advancements in LLMs necessitate a careful examination of their implications for both summative and formative assessment practices, particularly in light of findings that these models demonstrate advanced apparent conceptual understanding and problem-solving capabilities—even surpassing human performance.

### A. Summative assessment

A significant challenge for summative assessment in physics education arises with regard to the integrity of such assessment in certain contexts. A central concern lies in the authenticity of student work submitted in unsupervised or lightly supervised summative contexts, such as take-home exams, online quizzes, open-book assignments, project reports, lab write-ups, or preliminary science competition rounds completed at home. As LLMs increasingly outperform human students on tasks requiring conceptual understanding and problem-solving abilities, it becomes increasingly difficult to determine whether a student-submitted solution reflects genuine physics understanding or simply the "ability" to copy and paste a problem into an LLM and more or less uncritically replicate the output [1]. Although we found that LLM-generated solutions typically exhibit lower overall variance and greater internal consistency than student responses, this characteristic—while potentially indicative—does not reliably allow for the conclusive identification of AI-generated work. Educators may take this as signals of LLM use, but not as definitive evidence, thereby limiting enforcement options. Instead, the observations of potential AI use by students should serve as a prompt for the necessity to explicitly address the use of LLMs with students—discussing not only their potentials and pitfalls, but also appropriate strategies for their effective use.

A second important challenge concerns equity and fairness in assessment. Access to high-performing proprietary LLMs is often restricted by paywalls or usage limitations, leading to disparities between students based on institutional support and personal or familial financial resources. These inequalities are particularly pronounced in competitive academic settings, such as science competitions, where early-stage success (e.g., in homework-based qualification rounds) may unintentionally depend on whether a student has access to, or chooses to use, an advanced LLM. From a motivational standpoint, such use of LLMs can also undermine students' sense of competence. When success is attributed to external tools rather than personal effort or understanding, the experience may fail to produce meaningful learning gains or long-term academic confidence. In such cases, even high-scoring outcomes may yield little educational benefits. These concerns point to the need for critical reflection on how LLM access and usage intersect with the intended goals of summative assessment. Educators (and science competition organizers alike) should consider how to design assessment formats that uphold fairness, promote genuine engagement, and avoid reinforcing digital access gaps.

Given these challenges, there is a pressing need to reconsider the design of summative assessments. Traditional formats such as open-book or take-home exams—which have always posed monitoring challenges—are particularly vulnerable to unregulated use of LLMs. To address this, educators may consider a shift toward assessment formats that are less susceptible to AI assistance, including in-person written or oral exams, which better ensure independent student work (see also Ref. [3]). If such formats are not feasible, alternative strategies should focus on designing tasks that current LLMs still struggle with. One promising approach is to incorporate problems that require interpreting non-textual information, such as diagrams, graphs, or experimental setups—areas in which even state-of-the-art LLMs continue to show limitations [78–80]. Another approach is to rely more heavily on ill-defined problems, which current LLMs also still seem to struggle with [19], however, further research is necessary. In any case, the emphasis of summative assessment should shift toward assessing students' modeling processes, reasoning strategies, and justifications, rather than focusing solely on arriving at a final answer. Another option might be to explicitly incorporate AI into assessment settings—for example, by allowing students to interact with LLMs while monitoring the dialogue and considering it as part of the evaluation process. However, such formats may increase grading complexity and thereby the workload of educators, making feasibility a key concern—especially in large student courses. Overall, most of these adjustments to summative assessment practices should be viewed as temporary solutions tailored to the current limitations of LLMs only. The rapid pace of AI development suggests that future models may soon be capable of handling visual information and ill-defined problems more effectively. As such, no single redesign strategy beyond insisting on in-person and oral exams will offer a long-term panacea. Thus, summative assessment practices must continuously evolve in tandem with technological advancements to preserve their validity, fairness, and pedagogical value.

**B. Formative assessment**

In contrast to the challenges posed in summative contexts, LLMs offer several promising affordances for formative assessment, particularly in supporting self-directed and unsupervised learning. As demonstrated in this study, LLMs can generate immediate, relatively consistent, and high-quality example solutions to advanced physics problems. A particularly meaningful use arises when students first attempt to solve a problem on their own and then use an LLM-generated solution for comparison and self-evaluation. This process encourages reflection on alternative strategies, identification of errors, and a deeper understanding of underlying concepts—thereby aligning well with the goals of formative assessment. Moreover, our findings, along with prior work, suggest that advanced prompt engineering is often unnecessary to elicit strong performance from LLMs regarding physics problem solving, lowering the barrier for integration into everyday learning routines.

Nonetheless, prior work has highlighted the risk of students accepting AI-generated responses without reflection [1]. To address this, educators should help students understand the strengths and weaknesses of LLM-generated solutions, which falls under the term of AI literacy [103,104]. Clear guidance on how to engage productively with such tools is essential to fostering responsible use. Crucially, evaluating whether an LLM-generated solution is acceptable or accurate, particularly in domains like physics, requires a non-negligible level of subject-specific expertise. Without a solid grounding in the relevant physics concepts, students may struggle to identify subtle errors or misconceptions in the AI's reasoning, potentially reinforcing misunderstandings rather than correcting them

LLMs also show potential for feedback generation regarding problem solving (e.g., [105–107]), particularly as their problem-solving capabilities appear to be related to their capability to assess student responses [13]. However, such LLM-based feedback should serve as a complement to, not a replacement for, teacher guidance, as several studies have already highlighted potential negative learning outcomes when AI is used in place of traditional instructional methods [108,109]. Moreover, the rapid pace of LLM development raises concerns about the long-term consistency and reliability of such feedback (e.g., due to temporal variability). Developers, educators, and students must remain mindful of these limitations and recognize that LLM-generated feedback—while often helpful—is not always accurate, contextually appropriate, or educationally sufficient.

## VII. CONCLUSION

The findings of this study indicate that the tested LLMs (*GPT-4o* and *o1-preview*) demonstrate advanced apparent problem-solving capabilities on Olympiad-type physics problems, on average exceeding the performance of actual Physics Olympiad participants. These results prompt a critical rethinking of instructional priorities and assessment strategies in physics education. As LLMs now perform exceptionally well not only on conceptual physics tasks but also on well-defined physics problems as typically encountered in physics instruction, educators must reconsider what remains most essential to teach in a world where such AI tools are widely available. Simply prohibiting LLM use is neither realistic nor pedagogically sound. Instead, physics education—particularly assessment—must adapt: both to protect the validity of summative evaluations and to leverage the potential of LLMs for formative purposes. Ultimately, students must learn to engage with these technologies reflectively and responsibly, as they are likely to become integral to future professions in science and even beyond.

**Author contribution statement (CrediT):**

**PT:** conceptualization, investigation, methodology, software, data curation, formal analysis, visualization, project administration, writing (original draft & finalization)

**HM:** conceptualization, investigation, data curation, writing (review & editing)

**FK:** conceptualization, writing (review & editing)

**BK:** data curation, writing (review & editing)

**SP:** conceptualization, funding acquisition, resources, writing (review & editing)

**PW:** conceptualization, funding acquisition, resources, writing (review & editing)

**Acknowledgements:**

This work was supported by the Klaus-Tschira-Stiftung (project *WasP*) under grant number 00.001.2023. The German Physics Olympiad is supported by the Federal Ministry of Education and Research under grant number 16SJW2553.

**Language editing:**

In the process of refining the manuscript, language editing was conducted with the assistance of ChatGPT, a language model developed by OpenAI. ChatGPT was employed to enhance the clarity, coherence, and style of the text while maintaining the integrity of the original content.

**Data availability statement:**

The LLM-generated solutions analyzed in this study, which are in German, are available in the OSF repository associated with this article. The repository is available at Ref. [110].

# Supplemental Part A: Translated Physics Olympiad mechanics problems used in the study

Tasks and solutions published with permission of the German Physics Olympiad.

### Task 1  Helicopter on Mars [HEL]                                                         (10 Pts.)
*(1st stage, IPhO 2024)*

On 19 April 2021, a human-developed aircraft took off from Mars for the first time. The helicopter *Ingenuity* has since completed numerous flights in the Martian atmosphere. The two counter-rotating rotors of the aircraft have a diameter of 1,2 m and the total mass of the helicopter is around 1,8 kg.

When developing *Ingenuity* on Earth, the conditions in the Martian atmosphere had to be taken into account. On Earth, whose atmosphere has a density of 1,2 kg m$^{-3}$ near the ground, the helicopter took off at 500 rotor revolutions per minute. In contrast, the density of the Martian atmosphere near the ground is only around 0,020 kg m$^{-3}$. Mars has a diameter of about $6{,}8 \cdot 10^3$ km and a mass of about $6{,}4 \cdot 10^{23}$ kg.

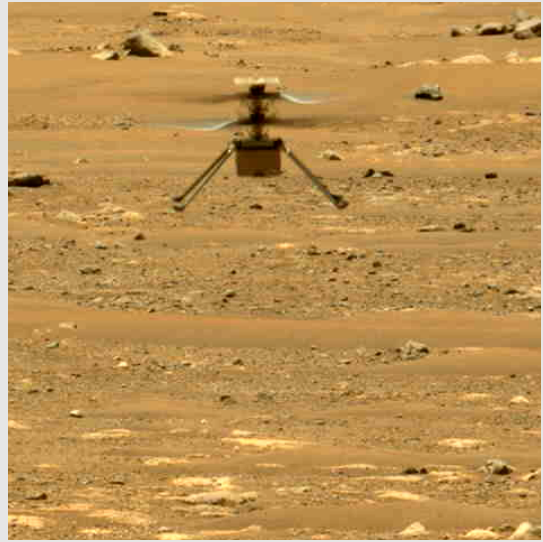

Abb. 1. *Ingenuity flying on Mars (Picture: NASA, en.wikipedia.org).*

Use the given data to determine the gravitational acceleration on Mars. Estimate the minimum number of revolutions per minute at which the rotors of *Ingenuity* must rotate on Mars for the helicopter to take off.

**Solution**

The helicopter's rotors push air, or more precisely gas from the respective atmosphere, downwards. The mean vertical velocity $\bar{v}$ of the gas deflected in this way is proportional to the rotational speed of the rotor blades. The following therefore applies

$$\bar{v} \sim \omega\, r \sim f\, r\,, \tag{1.1}$$

where $\omega$ and $f$ indicate the angular velocity and the frequency of the rotor movement and $r$ the rotor radius.

The air mass pushed downwards per time $\Delta m/\Delta t$ is in turn proportional to this speed $\bar{v}$, to the rotor area $A \sim r^2$ and to the density $\rho$ of the air:

$$\frac{\Delta m}{\Delta t} \sim \rho\, A\, \bar{v} \sim \rho\, f\, r^3\,. \tag{1.2}$$

The mass pushed downward therefore undergoes a change in momentum per time, which is given by

$$\frac{\Delta p}{\Delta t} = \frac{\Delta m}{\Delta t}\, \bar{v} \sim \rho\, f^2\, r^4\,. \tag{1.3}$$

According to Newton's second and third laws, this change in momentum causes an upward force of $\Delta p/\Delta t$ on the helicopter. This force must compensate the weight $mg$ of the helicopter so that the helicopter can take off. The following must therefore hold:

$$\boxed{mg = \frac{\Delta p}{\Delta t} = \kappa \rho f^2 r^4}. \tag{1.4}$$

We assume that the proportionality constant $\kappa$ in equation (1.4) is the same on Earth and Mars. This is approximately justified, since the relevant geometry and properties of the environment have been captured in equations (1.1) and (1.2).

Use the indices "E" and "M" to denote the corresponding variables on Earth and Mars. From (1.4), one obtains:

$$\frac{m}{\kappa r^4} = \frac{\rho_E f_E^2}{g_E} = \frac{\rho_M f_M^2}{g_M}. \tag{1.5}$$

The gravitational acceleration $g_M$ on Mars can be determined with the given diameter $d$ and the mass $M$ of the planet using Newton's law of gravity as

$$\boxed{g_M = G \frac{M}{d^2/4} \approx 3{,}7\,\mathrm{m\,s^{-2}}}, \tag{1.6}$$

with gravitational constant $G = 6{,}674 \cdot 10^{-11}\,\mathrm{m^3\,kg^{-1}\,s^{-2}}$.

This can be used to determine the approximate rotational frequency required on Mars:

$$\boxed{f_M = f_E \sqrt{\frac{g_M}{g_E} \frac{\rho_E}{\rho_M}} \approx 4{,}8\,f_E = 2{,}4 \cdot 10^3\,\mathrm{min^{-1}}}. \tag{1.7}$$

The rotor on Mars must therefore rotate at about 2400 revolutions per minute[1].

*Note:* The idea for this task comes from the article Blanco, P. (2021). *Rotorcraft RPM on Mars.* arxiv. https://arxiv.org/abs/2105.00797.

| Scoring scheme - Helicopter on Mars [HEL] | | points |
|---|---|---|
| 1 | Formulation of the idea that the rotor pushes gas downwards | 1.0 |
| | Propose a proportional relationship for the mass of gas pushed down per unit of time by the rotor (1.2) | 1.0 |
| | Specify the change in momentum per unit of time (i.e., the upward force) (1.3) | 1.0 |
| | Application of force equilibrium (1.4) | 1.0 |
| | Use of proportionality to establish a relationship between relevant quantities on Earth and Mars (1.5) | 2.0 |
| | Determine the gravitational acceleration on Mars (1.6) | 2.0 |
| | Determine the result for the rotor frequency (1.7) | 2.0 |
| | | **10.0** |

---

[1]This value corresponds very well to the actual rotation frequency of the rotors on Mars (see e. g. https://en.wikipedia.org/wiki/Ingenuity_(helicopter))

**IPN** Leibniz-Institut für die Pädagogik der Naturwissenschaften und Mathematik

## Task 2 Capsizing Iceberg [ICE] (10 Pts.)

*(1st stage, IPhO 2022)*

Tabular icebergs have a relatively flat top and steep edges. They are formed by ice shelves breaking off and can be quiet large. Just like other icebergs, most of the ice of a floating tabular iceberg is not above the water but below it.

In the following, consider an approximately cuboid iceberg of height $H$, length $L$ and width $W$. Use the approximate values $\rho_{\text{ice}} = 0{,}9 \cdot 10^3\,\text{kg}\,\text{m}^{-3}$ and $\rho_{\text{water}} = 1{,}0 \cdot 10^3\,\text{kg}\,\text{m}^{-3}$ for the density of ice and seawater.

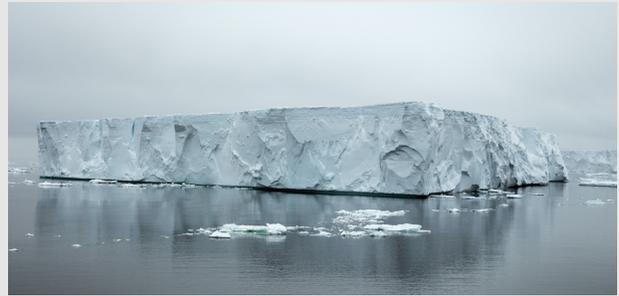

Abb. 2. *Photo of a tabular iceberg (by Andrew Shiva, CC BY-SA 4.0).*

a) Determine which proportion of the height $H$ of the iceberg is below the water surface.

When an ice shelf breaks off, relatively narrow tabular icebergs may form, where the width $W$ is smaller than the height $H$. Such icebergs have the potential to tip over onto their sides.

b) Show that it is energetically more favorable for a narrow iceberg to tip onto its side. Determine the ratio of width to height that maximizes the energy released during this capsizing process. Assume the condition $L > H$ holds.

The released energy can cause high waves, which can also be dangerous for nearby ships.

The photo on the right shows the passenger ship Fram in front of an iceberg. The ship has a length of about 114 m.

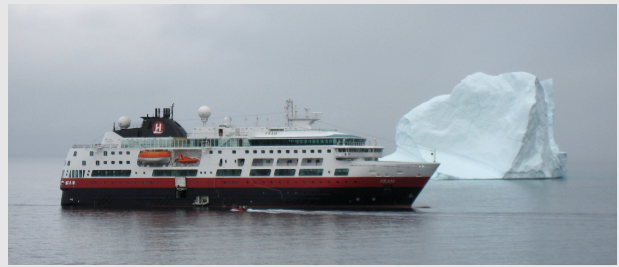

Abb. 3. *Photo of a ship in front of an iceberg (by Kim Hansen, CC BY-SA 4.0).*

c) Estimate the energy that would be released when the iceberg tips over. For simplicity, assume that the photo in Figure 3 was taken from a great distance and that the iceberg has approximately the shape of a tabular iceberg with the same length and width. Calculate the mass of TNT that would have to be detonated to release the same amount of energy.

**Solution**

a) The iceberg floats if its weight is equal to the buoyant force acting on it. According to Archimedes' principle, the buoyant force is equal to the weight of the water displaced by the iceberg. Let the immersion depth of the cuboid-shaped iceberg be denoted by $D$. The equilibrium of forces is then expressed by

$$\rho_{\text{ice}}\, H\, L\, W\, g = \rho_{\text{water}}\, D\, L\, W\, g, \quad \text{and therefore} \quad \boxed{D = \frac{\rho_{\text{ice}}}{\rho_{\text{water}}}\, H \approx 0{,}9\, H}. \tag{2.1}$$

Thus, about 90 % of the height of the iceberg is under water.

IPN Leibniz-Institut für die Pädagogik der Naturwissenschaften und Mathematik

b) The following sketches show the positions of the floating tabular iceberg before and after capsizing.

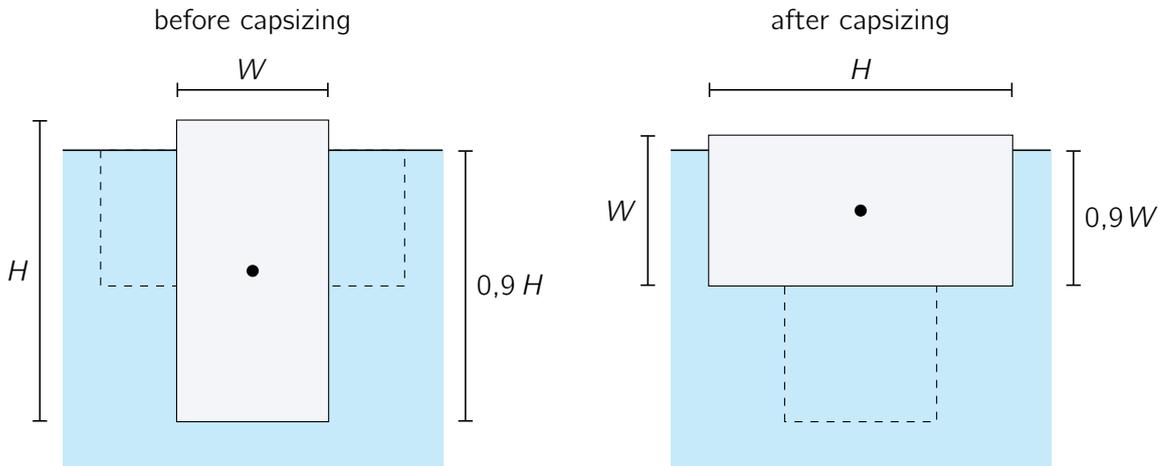

Abb. 4. *Sketches of the tabular iceberg before and after capsizing. The dashed lines indicate the volumes of water displaced during capsizing.*

To tip the iceberg, its center of gravity must be raised by a height of $\frac{4}{10}H - \frac{4}{10}W$. Achieving this requires supplying the iceberg with potential energy, given by

$$\Delta E_{\text{ice}} = \rho_{\text{ice}} H L W g \frac{4}{10}(H - W). \tag{2.2}$$

At the same time, the potential energy of the water decreases, as water closer to the surface now fills the volume previously occupied by the iceberg. From Figure 4, the displaced water volume is given as $0{,}9(H - W)LW$. The center of gravity of this water volume is initially located at a depth of $\frac{0{,}9}{2}W$ below the water surface. After the iceberg tips, the center of gravity shifts to a depth of $0{,}9W + \frac{0{,}9}{2}(H - W)$.

The difference in the potential energy of the water is therefore

$$\Delta E_{\text{water}} = \rho_{\text{water}} H L W g \frac{0{,}9^2}{2}(H - W). \tag{2.3}$$

Hence, the total energy released is

$$\boxed{\Delta E = \Delta E_{\text{water}} - \Delta E_{\text{ice}} = H L W g (H - W) \left(\frac{0{,}9^2}{2} \rho_{\text{water}} - \frac{4}{10}\rho_{\text{ice}}\right) > 0}. \tag{2.4}$$

Equation (2.4) can be rewritten with $k := g\left(\frac{0{,}9^2}{2}\rho_{\text{water}} - \frac{4}{10}\rho_{\text{ice}}\right) \approx 450\,\text{J}\,\text{m}^{-4}$ as

$$\Delta E = k L H^3 \frac{W}{H}\left(1 - \frac{W}{H}\right) = k L H^3 \left\{\frac{1}{4} - \left(\frac{W}{H} - \frac{1}{2}\right)^2\right\}. \tag{2.5}$$

The expression within the curly brackets is maximized when the quadratic term equals zero, i.e., for $W = \frac{1}{2}H$. The energy released is therefore maximized when the width of the tabular iceberg is half its height. This solution can, for example, also be obtained graphically or through an extreme value analysis.

c) To provide a rough estimate, the dimensions of the iceberg in the photo can be compared with the length of the ship also visible in the photo. Denoting the ship's length as $\ell = 114\,\text{m}$, the width of the iceberg is estimated from the photo as $W = L \approx 0{,}45\,\ell \approx 51\,\text{m}$, while the height is approximated as $H \approx 10 \cdot 0{,}30\,\ell \approx 344\,\text{m}$. Using these values in Equation (2.5), the energy released is calculated as

$$\boxed{\Delta E \approx 1{,}2 \cdot 10^{11}\,\text{J}}. \tag{2.6}$$

One kilogram of TNT releases approximately $4{,}2 \cdot 10^6$ J of energy upon explosion (see, for example, the entry on TNT equivalent on Wikipedia). The mass $m_\text{TNT}$ of TNT required to release the same amount of energy can be calculated as:

$$\boxed{m_\text{TNT} \approx \frac{\Delta E}{4{,}2 \cdot 10^6\,\text{J}\,\text{kg}^{-1}} \approx 2{,}9 \cdot 10^4\,\text{kg}}, \tag{2.7}$$

which corresponds to approximately 29 tons.

If participants use other plausible values for their estimation (e.g., for the visible dimensions of the iceberg), these should also be regarded as valid and accurate.

*Note:* The idea of this task is drawn from the article Marshall, R. (2015). Capsizing icebergs: an exercise in the application of the principle of the conservation of energy with a very surprising result. *Physics Education*, **50**(3).

| **Evaluation - Capsizing Iceberg [ICE]** | | points |
|---|---|---|
| 2.a) | Using the equilibrium of forces and Archimedes' principle | 1.0 |
| | Determine the percentage below the water surface (2.1) | 1.0 |
| 2.b) | Determine the change in the potential energy of the iceberg (2.2) | 1.0 |
| | Determine the change in the potential energy of the water (2.3) | 1.0 |
| | Determine the total energy released (2.4) or (2.5) | 1.0 |
| | Determine the condition $W = \frac{1}{2}H$. | 2.0 |
| 2.c) | Estimate the relevant quantities from the photo | 1.0 |
| | Calculate the energy released during tilting (2.6) | 1.0 |
| | Determine a matching TNT equivalent (2.7) | 1.0 |
| | | **10.0** |

## Task 3 Fall on Exoplanet [EXO] (5 Pts.)

*(2nd stage, IPhO 2022)*

On the surface of an extrasolar planet – or exoplanet for short - the fall time of a body from a small height $h$ is exactly twice as long as on Earth, neglecting frictional effects.

Which of the following statements is compatible with this description, assuming the exoplanet to be a spherically symmetrical object?

The exoplanet has ...

A ... half the mass and twice the radius of the Earth.

B ... exactly the mass and four times the radius of the Earth.

C ... twice the mass and twice the radius of the Earth.

D ... four times the mass and four times the radius of the Earth.

**Solution**

Calculations and explanations: The fall time for a free fall from height $h$ on a planet's surface is determined from $h = \frac{1}{2} g t^2$:

$$t = \sqrt{\frac{2h}{g}}. \tag{3.1}$$

The gravitational acceleration $g$ on the planet can be expressed with the help of Newton's gravitational law by

$$g = \frac{G M}{R^2}, \tag{3.2}$$

where $M$ is the mass of the planet, $R$ its radius and $G$ the gravitational constant. For the fall time, this results in

$$t = \sqrt{\frac{2 h R^2}{G M}}. \tag{3.3}$$

If this fall time is to be twice as long for a fixed $h$, then $\frac{R_{\text{Exo}}^2}{M_{\text{Exo}}} = 4 \cdot \frac{R_{\text{Earth}}^2}{M_{\text{Earth}}}$ must apply. This is only true for answer option D.

Correct answer: D

| Evaluation - Fall on Exoplanet [EXO] | | points |
|---|---|---|
| 3 | Specify the fall time during free fall | 1.0 |
| | Express the gravitational acceleration in terms of mass and radius | 1.0 |
| | Derive the correct dependence of the time of fall on $R$ and $M$ | 1.0 |
| | Specify the correct solution | 2.0 |
| | | **5.0** |

## Task 4 Rocket Launches and Satellites [ROC] (20 Pts.)

*(2nd stage, IPhO 2023)*

The number of rocket launches has increased significantly in recent years, with more than 140 launches aimed at reaching Earth orbit in 2021 alone. During the launch phase, rockets and their payloads are subjected to extreme stresses, among which aerodynamic stress, caused by atmospheric friction, plays a pivotal role.

As a simplified model, consider a rocket with a cone-shaped tip characterized by a diameter $d$ and an opening angle $\alpha$ at its apex. The rocket travels at a speed $v$ through the atmosphere, which, at its current altitude, has a density of $\rho_{atm}$. You can assume that the movement of the air molecules in the atmosphere is negligible compared to the speed of the rocket. The rocket experiences a frictional force due to elastic collisions between its tip and the air molecules.

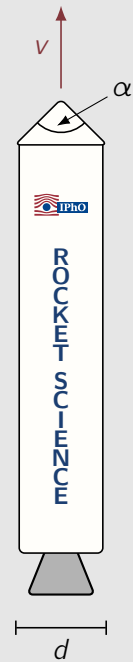

4.a) Derive an expression for the frictional force acting on the rocket as a function of the parameters $d$, $\alpha$, $v$ and $\rho_{atm}$. Calculate the frictional force for the following values: $d = 3{,}7\,\text{m}$, $\alpha = 90°$, $v = 2{,}0\,\text{km}\,\text{s}^{-1}$, and $\rho_{atm} = 1{,}0 \cdot 10^{-3}\,\text{kg}\,\text{m}^{-3}$. *(4.0 Pts.)*

The frictional force acting on a rocket changes during its flight. The following figures show the speed of a rocket after launch as a function of height above the ground (left) and the atmospheric pressure as a function of height above the ground (right). For simplicity, it is assumed that the temperature of the atmosphere remains constant.

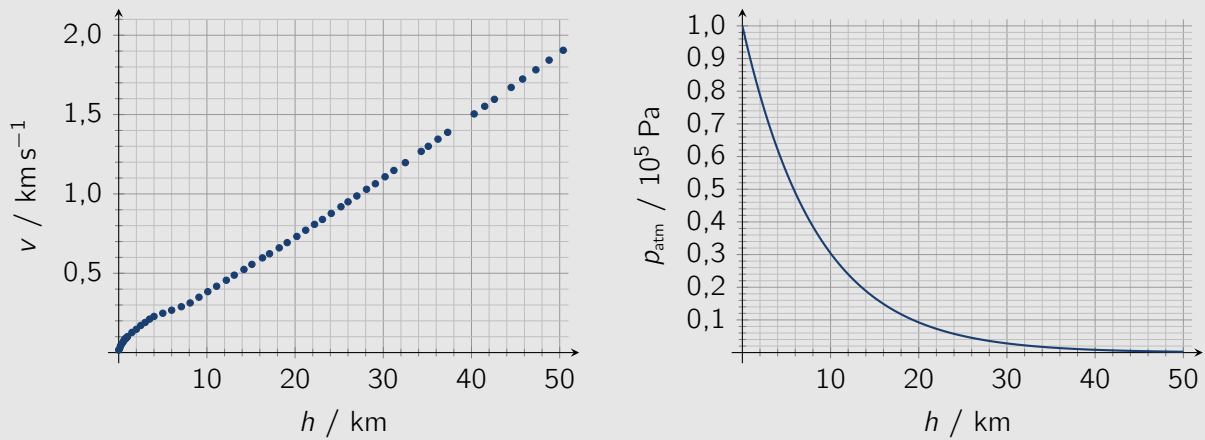

Abb. 5. *Velocity $v$ of the rocket (left) and air pressure $p_{atm}$ of the atmosphere (right) as a function of the height $h$ above the ground.*

4.b) Using the data from the graphs, estimate the height above the ground at which the frictional force acting on the rocket reaches its maximum. *(6.0 Pts.)*

This critical point during a launch, known as Max Q, represents the location and time when the rocket experiences maximum aerodynamic stress.

![IPN] **Leibniz-Institut für die Pädagogik der Naturwissenschaften und Mathematik**

To place satellites into orbit around Earth, the rocket must continue to accelerate. Let the mass of Earth be denoted as $m_E = 5{,}97 \cdot 10^{24}$ kg and the radius of the Earth as $R_E = 6{,}37 \cdot 10^6$ m.

4.c) Determine the speed to which the rocket must accelerate before shutting off its engines in order to achieve a stable near-Earth orbit outside the atmosphere, without crashing back to the surface. Also, calculate the orbital period for this orbit. *(3.0 Pts.)*

4.d) Determine the minimum speed to which the rocket must accelerate before shutting off its engines to completely escape Earth's gravitational influence. Calculate the ratio of this speed to the speed determined in the previous task. *(3.0 Pts.)*

Now consider a satellite orbiting the Sun at a radius approximately equal to the Earth's average orbital radius of $1{,}5 \cdot 10^{11}$ m. The satellite is located far from Earth and other celestial bodies. The mass of the Sun is approximately $m_S = 1{,}99 \cdot 10^{30}$ kg, and the Sun's radius can be assumed to be negligible compared to the Earth's orbital radius. Suddenly, the satellite comes to a complete stop relative to the Sun.

4.e) Estimate the time it would take for the satellite to crash into the Sun. Depending on the chosen approach, Kepler's laws may prove helpful. *(4.0 Pts.)*

**Solution**

4.a) Calculations and explanations

In the reference frame of the rocket, the air molecules collide head-on with the tip of the rocket at a speed of $v$. During these collisions, which are assumed to be elastic, the molecules are deflected by an angle $\alpha$ relative to their original direction of motion, as illustrated in the figure on the right. During this deflection, the magnitude of their velocity remains unchanged. The momentum transferred to the rocket by an air molecule of mass $m_{\text{molecule}}$ in the direction of the molecule's initial motion is given by $\Delta p_{\text{molecule}}$.

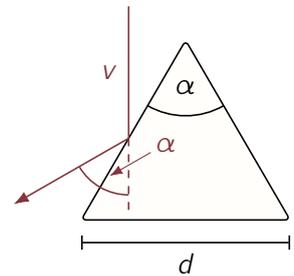

$$\Delta p_{\text{molecule}} = m_{\text{molecule}}\, v\, (1 - \cos\alpha)\,. \tag{4.1}$$

During a small time interval $\Delta t$, the number of air molecules colliding with the rocket is given by

$$\Delta n = A\, v\, \Delta t\, \frac{\rho_{\text{atm}}}{m_{\text{molecule}}} = \frac{\pi\, d^2\, v\, \rho_{\text{atm}}}{4\, m_{\text{molecule}}} \Delta t\,, \tag{4.2}$$

where $A = \pi d^2/4$ represents the cross-sectional area of the rocket. Consequently, the total momentum transferred to the rocket by the air molecules per time unit is

$$\boxed{F = \frac{\Delta p}{\Delta t} = \frac{\Delta n\, \Delta p_{\text{molecule}}}{\Delta t} = \frac{\pi}{4} d^2\, v^2\, \rho_{\text{atm}}\, (1 - \cos\alpha)}\,. \tag{4.3}$$

According to Newton's second law, the rate of change of momentum over time corresponds directly to the frictional force $F$ acting on the rocket. Using the provided values of $d = 3{,}7$ m, $\alpha = 90°$, $v = 2{,}0\,\text{km}\,\text{s}^{-1}$ and $\rho_{\text{atm}} = 1{,}0 \cdot 10^{-3}\,\text{kg}\,\text{m}^{-3}$, the frictional force is calculated as

$$\boxed{F \approx 4{,}3 \cdot 10^4\,\text{N}}\,. \tag{4.4}$$

4.b) Calculations and explanations

According to (4.3), the frictional force is proportional to the square of the rocket's speed and the density of the air. Assuming the atmospheric temperature remains constant, the ideal gas law implies that air density is proportional to air pressure. Consequently, the frictional force acting on the rocket is proportional to the product $v^2 \cdot p_{atm}$. All other factors in (4.3) depend solely on the rocket's geometry, which remains unchanged throughout the flight.

To identify the point at which the frictional force acting on the rocket reaches its maximum, it is sufficient to determine the maximum value of $v^2 \cdot p_{atm}$ as a function of the height $h$. This can be done graphically using the data provided in the graphs, as illustrated in the figure on the right.

From the graph, the height at which the maximum frictional force acts on the rocket is approximately

$$\boxed{h_{\text{Max Q}} \approx (15 \pm 1)\,\text{km}}. \quad (4.5)$$

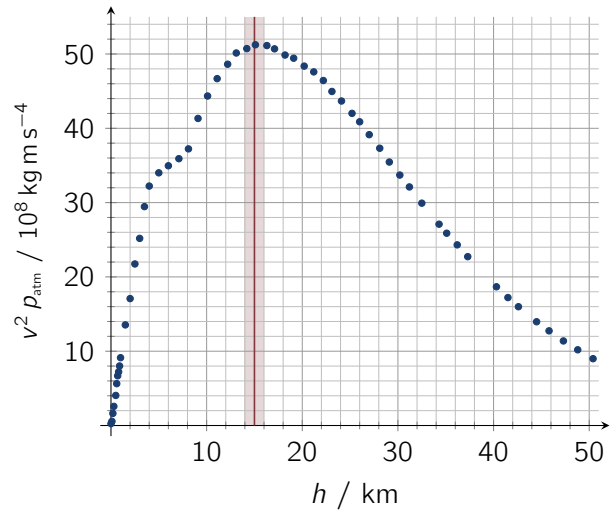

Abb. 6. *Product $v^2 \cdot p_{atm}$ at rocket launch as a function of the height $h$ above the ground*.

*Note:* For the estimation in the exam, it is sufficient to evaluate the data point by point and use it to approximately determine the height.

Alternatively, the height can be estimated using the following reasoning: beyond the first approximately 7 km, the speed $v$ is, to a good approximation, proportional to the height $h$. This implies that the frictional force is approximately proportional to $h^2 e^{-h/h_{scale}}$, where the scale height $h_{scale}$ is 8,4 km, as determined from the pressure graph. By setting the derivative of this function to zero, the height at which the frictional force is maximized can be estimated as twice the scale height, i.e., approximately 16,8 km.

4.c) Calculations and explanations

To achieve an orbit around Earth outside the atmosphere, allowing for nearly frictionless motion at a speed of $v$, the centripetal force acting on the rocket must be entirely provided by gravitational force. Denoting the mass of the rocket as $m$ and the radius of the orbit as $R$, the following condition must be satisfied:

$$F_{\text{centripetal}} = \frac{m\,v^2}{R} = G\,\frac{m\,m_E}{R^2} = F_{\text{gravity}}. \quad (4.6)$$

Where $G = 6{,}674 \cdot 10^{-11}\,\text{m}^3\,\text{kg}^{-1}\,\text{s}^{-2}$ is the gravitational constant and $m_E = 5{,}97 \cdot 10^{24}\,\text{kg}$ is the mass of the Earth.

The Earth's atmosphere is extremely thin compared to the Earth's diameter (as illustrated in the atmospheric pressure graph provided in the task). Therefore, for a near-Earth orbit, the orbital radius can be approximated as $R_E = 6{,}37 \cdot 10^6\,\text{m}$. By rearranging (4.6), the required orbital speed $v_1$ can be calculated as follows

$$\boxed{v_1 = \sqrt{\frac{G\, m_E}{R_E}} = 7{,}91\,\mathrm{km\,s^{-1}}}. \tag{4.7}$$

This speed is commonly referred to as the first cosmic velocity. For slightly larger assumed orbital radii, the value decreases slightly. The corresponding orbital period is

$$\boxed{T_1 = \frac{2\pi R_E}{v_1} = 2\pi R_E \sqrt{\frac{R_E}{G\, m_E}} = 5{,}06\cdot 10^3\,\mathrm{s} \approx 84{,}3\,\mathrm{min}}. \tag{4.8}$$

**4.d)** Calculations and explanations

To escape Earth's gravitational influence, the rocket must achieve a speed that ensures it has zero velocity at an infinite distance from Earth. Otherwise, the rocket would fall back to Earth.

Since the total energy (the sum of the kinetic and potential energy) of the rocket remains constant when it is outside the atmosphere, and the potential energy approaches zero at a great distance from Earth, the rocket's initial kinetic energy near Earth must equal its potential energy in Earth's gravitational field. Therefore, the following relationship, based on the terms from the previous part of the task, must hold:

$$0 = E_{\mathrm{kin}}(R_E) + E_{\mathrm{pot}}(R_E) = \frac{1}{2}m v^2 - G\frac{m\, m_E}{R_E}\,. \tag{4.9}$$

Rearranging for the velocity, the so-called second cosmic velocity is given by

$$\boxed{v_2 = \sqrt{\frac{2G\, m_E}{R_E}} = \sqrt{2}\, v_1 = 11{,}1\,\mathrm{km\,s^{-1}}}. \tag{4.10}$$

The velocity is therefore greater by a factor of $\sqrt{2}$ compared to the velocity required to maintain a circular orbit.

**4.e)** Calculations and explanations

**Solution variant 1** - According to Kepler's third law, the ratio of the square of the orbital period to the cube of the semi-major axis is constant for all bodies orbiting the Sun on elliptical paths. This relationship is expressed as:

$$\frac{T^2}{a^3} = \mathrm{const.} = \frac{T_E^2}{r_E^3}\,, \tag{4.11}$$

where $T_E = 1{,}0\,\mathrm{a}$ represents the orbital period of the Earth around the Sun, and $r_E = 1{,}5\cdot 10^{11}\,\mathrm{m}$ is the distance between the Earth and the Sun, assuming a circular orbit. After deceleration, the satellite falls radially into the Sun. Its orbit can be approximated as an ellipse with a negligibly small minor axis. In this case, the focal point, which corresponds to the center of the Sun, coincides with the satellite's perihelion (closest point to the Sun). Consequently, the semi-major axis $a_{\mathrm{satellite}}$ of the orbit is half the Earth-Sun distance. Using

(4.11), the time $t$ until the satellite crashes into the Sun can be determined as follows:

$$t = \frac{T_{\text{satellite}}}{2} = \frac{T_E}{2}\left(\frac{a_{\text{satellite}}}{r_E}\right)^{\frac{3}{2}} = \frac{T_E}{4\sqrt{2}} \approx 0{,}177\,\text{a} \approx 65\,\text{days}. \qquad (4.12)$$

**Solution variants 2 & 3** - Starting from the law of conservation of energy[a], the satellite's velocity $v$ at any distance $r$ from the Sun during its fall can be determined. This way, the following holds, analogously to (4.9):

$$\frac{1}{2} m_{\text{satellite}}\, v^2 - G\, \frac{m_{\text{satellite}}\, m_S}{r} = -G\, \frac{m_{\text{satellite}}\, m_S}{r_E}, \qquad (4.13)$$

where $m_{\text{satellite}}$ is the satellite's mass, and $m_S$ is the Sun's mass. From this, the velocity during the fall is given by

$$v = -\dot{r} = \sqrt{\frac{2\,G\,m_S}{r_E}}\, \sqrt{\frac{r_E - r}{r}}\,. \qquad (4.14)$$

This differential equation can be integrated, ultimately yielding the same result as (4.12)[b]

Alternatively, the fall time can also be estimated numerically using (4.14). In this approach, the fall velocity is calculated at specific distances and used as the average velocity for each segment of the fall. Dividing the initial distance of the satellite from the Sun into ten equal segments yields an estimated fall time of about 57 days.

*Note:* This part of the task can also be found in the book: Geckeler, C. & Lind, G. (2002). *Physik zum Nachdenken: 100 Olympiade-Aufgaben mit Lösungen* (2nd ed.). Aulis-Verlag, Cologne.

---

[a]Alternatively, one could begin with the equation of force in the Sun's gravitational field. However, this approach requires integrating the equation of motion once to arrive at the conservation of energy

[b]Since the integration is nontrivial, a complete solution is unlikely to appear in an exam setting. For completeness, the integration proceeds as follows: Using (4.14), separating the variables leads to

$$-\sqrt{\frac{r}{r_E - r}}\, dr = \sqrt{\frac{2\,G\,m_S}{r_E}}\, dt \quad \text{or integrated} \quad -\int_{r_E}^{0} dr\, \sqrt{\frac{r}{r_E - r}} = \sqrt{\frac{2\,G\,m_S}{r_E}}\, t\,.$$

The definite integral on the left-hand side can be transformed by substitution. It is

$$\int_0^{r_E} dr\, \sqrt{\frac{r}{r_E - r}} \stackrel{x := r_E - r}{=} \int_0^{r_E} dx\, \sqrt{\frac{r_E}{x} - 1} \stackrel{y := \sqrt{\frac{r_E}{x} - 1}}{=} \int_0^{\infty} dy\, \frac{2\,r_E\,y}{(1 + y^2)^2}\, y\,.$$

The resulting integral can be solved by partial integration:

$$\int_0^{\infty} dy\, \frac{2\,r_E\,y}{(1 + y^2)^2}\, y = \left[-\frac{r_E}{1 + y^2}\, y\right]_0^{\infty} + \int_0^{\infty} dy\, \frac{r_E}{1 + y^2} = 0 + r_E\, [\arctan y]_0^{\infty} = r_E\, \frac{\pi}{2}\,.$$

Using $\frac{4\pi^2}{T_E^2}\, r_E = \frac{G\,m_S}{r_E^2}$, the orbital period of the Earth is given by $T_E = \frac{2\pi\, r_E^{2/3}}{\sqrt{G\,m_S}}$. This leads to the same satellite's fall time as in (4.12):

$$t = \frac{\pi\, r_E^{2/3}}{2\sqrt{2\,G\,m_S}} = \frac{T_E}{4\sqrt{2}} \approx 0{,}177\,\text{a} \approx 65\,\text{day}\,.$$

| Evaluation - Rocket Launches and Satellites [ROC] | | dots |
|---|---|---|
| 4.a) | Recognising the deflection of air molecules by angle $\alpha$ on impact | 0.5 |
| | Determining the change in momentum through collision of an air molecule on the rocket (4.1) | 1.0 |
| | Determining the number of air molecules hitting the rocket per time interval (4.2) | 0.5 |
| | Recognising that the rate of change of the momentum corresponds to the frictional force and giving an expression for the frictional force (4.3) | 1.0 |
| | Calculate the value for the frictional force (4.4) | 1.0 |
| 4.b) | Use of the proportionality of the frictional force to $v^2 \cdot \rho_{atm}$ | 1.0 |
| | Recognise that density can be assumed to be proportional to pressure | 1.0 |
| | Formulate an idea for determining the height corresponding to the maximum friction force | 1.0 |
| | Analysing the data to determine the height corresponding to the maximum friction force | 2.0 |
| | Determine the height with $14\,\text{km} \leq h_{\text{Max Q}} \leq 18\,\text{km}$ (4.5) | 1.0 |
| 4.c) | Equating centripetal and gravitational force (4.6) | 1.0 |
| | Transforming the force equation into the velocity (4.7) | 0.5 |
| | Using a radius close to the radius of the earth | 0.5 |
| | Calculate the value of the velocity in (4.7) | 0.5 |
| | Calculate the value of the orbital period in (4.8) | 0.5 |
| 4.d) | Using and specifying the energy theorem (4.9) | 1.0 |
| | Converting the energy theorem to velocity (4.10) | 0.5 |
| | Calculate the value of the velocity in (4.10) | 1.0 |
| | Determine the ratio of the two speeds | 0.5 |
| 4.e) | Recognise that the trajectory during a fall can be considered as a degenerate ellipse | 1.0 |
| | Using Kepler's 3rd law (4.11) | 1.0 |
| | Using the correct semi-major axis $a_{\text{satellite}} = r_E/2$ | 0.5 |
| | Recognise that fall time corresponds to half the orbital period | 0.5 |
| | Calculate the fall time (4.12) | 1.0 |
| | | **20.0** |

*Comment on the last part of the problem:* - If the solution is based on the law of conservation of energy, the points are awarded as follows: 1.0 Pt. for the correct formulation of the law of conservation of energy, 0.5 Pt. for the conversion to velocity (4.14), 0.5 Pt. for an idea for the evaluation, 1.0 Pt. for the evaluation, 1.0 Pt. for the (approximate) calculation of the time of fall, whereby all plausible values should be scored.

### Task 5 Insect Hunting [INS] (5.0 Pts.)

(3rd stage, IPhO 2023)

A bat hunting for food flies at a speed of $25\,\text{km}\,\text{h}^{-1}$ as it hunts an insect. The bat emits a sound at a frequency of $40{,}0\,\text{kHz}$ and detects an echo with a frequency of $40{,}4\,\text{kHz}$.

Determine whether the bat is moving towards the insect or away from it, and calculate the speed at which the bat is approaching the insect or the insect is receding from the bat.

You can assume a speed of sound in air of $343\,\text{m}\,\text{s}^{-1}$.

**Solution**

From the bat's perspective, the reflected signal received has a higher frequency than the emitted signal. This indicates that the bat is approaching the insect.

Since both the transmitter and receiver are in motion, the Doppler effect formula applies to the frequency $f'$ perceived by the receiver:

$$f' = f\,\frac{1 - \frac{v_R}{c_{\text{air}}}}{1 - \frac{v_T}{c_{\text{air}}}} = f\,\frac{c_{\text{air}} - v_R}{c_{\text{air}} - v_T}\,. \tag{5.1}$$

Here, $f$ represents the frequency of the signal emitted by the stationary transmitter, while $v_T$ and $v_R$ denote the velocities of the transmitter and receiver, respectively, relative to the stationary air. The signs of the velocities are chosen so that they have a positive sign in the direction of the movement of the transmitter.

Now, let $v_B$ represent the speed of the bat, and $v_I$ the speed of the insect, both measured relative to the stationary air. The bat emits a signal at a frequency $f$, which the moving insect perceives as $f'$. The insect reflects this signal, and the bat subsequently detects the reflected signal at a frequency

$$f'' = f \cdot \frac{1 + \frac{v_B}{c_{\text{air}}}}{1 + \frac{v_I}{c_{\text{air}}}} \cdot \frac{1 - \frac{v_I}{c_{\text{air}}}}{1 - \frac{v_B}{c_{\text{air}}}} = f \cdot \frac{c_{\text{air}} + v_B}{c_{\text{air}} + v_I} \cdot \frac{c_{\text{air}} - v_I}{c_{\text{air}} - v_B} \stackrel{!}{=} 40{,}4\,\text{kHz}\,. \tag{5.2}$$

Solving the equation above for the insect's speed $v_I$, and substituting the given values, results in

$$v_I = c_{\text{air}}\,\frac{c_{\text{air}}(f - f'') + v_B(f + f'')}{c_{\text{air}}(f + f'') + v_B(f - f'')} \approx 5{,}3\,\text{m}\,\text{s}^{-1} \approx 19\,\text{km}\,\text{h}^{-1}\,. \tag{5.3}$$

Thus, the insect moves away from the bat at this speed, while the bat approaches the insect at a speed given by

$$\boxed{\Delta v = v_B - v_I \approx 1{,}7\,\text{m}\,\text{s}^{-1} \approx 6\,\text{km}\,\text{h}^{-1}}\,. \tag{5.4}$$

| Scoring scheme - Insect Hunting [INS] | | points |
|---|---|---|
| 5 | Recognize and reason that the bat is approaching the insect | 1.0 |
| | Specify the Doppler-shifted frequency with a moving transmitter, see (5.1) | 1.0 |
| | Specify the Doppler-shifted frequency with a moving receiver, see (5.1) | 1.0 |
| | Double usage of the Doppler shift (5.2) | 1.0 |
| | Solve for the speed of the insect (5.3) | 0.5 |
| | Calculate the speed at which the bat is approaching the insect (5.4) | 0.5 |
| | | **5.0** |

### Task 6 Sled Pulling [SLE] (4.0 Pts.)

*(4th stage, IPhO 2023)*

> A child pulls a sled of mass $m$ at a constant speed up an inclined plane that is angled $\alpha$ relative to the horizontal. The pulling rope forms an angle $\beta$ with the inclined plane. The coefficient of sliding friction between the sled and the snow is $\mu$.
>
> Determine an expression for the angle $\beta$ at which the pulling force exerted by the child is minimized. Provide an expression for this minimum pulling force.

**Solution**

Let $\vec{P}$ represent the pulling force on the rope, and $\vec{W}$, $\vec{N}$ and $\vec{F}$ represent the sled's weight, the normal force exerted by the ground on the sled (perpendicular to the plane), and the frictional force (acting along the plane), respectively.

When the sled moves at a constant speed, these forces must be in equilibrium, meaning their components along and perpendicular to the plane must sum to zero:

$$0 \stackrel{!}{=} P \cos\beta - F - W \sin\alpha \qquad \text{as well as} \qquad 0 \stackrel{!}{=} P \sin\beta + N - W \cos\alpha. \tag{6.1}$$

For the absolute values of the forces, the following holds: $F = \mu N$ and $W = mg$.

Solving both equations in (6.1) for $N$, equating the resulting equations, and finally solving for the pulling force $P$ yields:

$$P = \frac{\sin\alpha + \mu \cos\alpha}{\cos\beta + \mu \sin\beta} \, mg. \tag{6.2}$$

The minimum value of $P$ as a function of $\beta$ occurs when the denominator is maximized. Setting the derivative $\partial(\cos\beta + \mu \sin\beta)/\partial\beta = -\sin\beta + \mu \cos\beta$ to zero gives the angle at which the pulling force is minimized.

$$\boxed{\beta_{\min} = \arctan\mu}. \tag{6.3}$$

Here, $\partial^2(\cos\beta + \mu \sin\beta)/\partial\beta^2 = -\cos\beta - \mu \sin\beta < 0$, confirming that the denominator reaches a maximum. Using the trigonometric identities

$$\cos\beta = \frac{1}{\sqrt{1+\tan^2\beta}} \qquad \text{as well as} \qquad \sin\beta = \frac{\tan\beta}{\sqrt{1+\tan^2\beta}}, \tag{6.4}$$

the minimum pulling force at this angle is:

$$\boxed{P_{\min} = \frac{\sin\alpha + \mu \cos\alpha}{\cos\beta_{\min} + \mu \sin\beta_{\min}} \, mg = \frac{\sin\alpha + \mu \cos\alpha}{\sqrt{1+\mu^2}} \, mg}. \tag{6.5}$$

| **Evaluation - Sled Pulling [SLE]** | | dots |
|---|---|---|
| 6 | Consider the relevant forces | 0.5 |
| | Recognising and setting up the equilibria of forces (6.1) | 1.0 |
| | Expressing the frictional force by normal force | 0.5 |
| | Specify the pulling force (6.2) | 1.0 |
| | Determine the angle for minimum force (6.3) | 0.5 |
| | Give an expression for minimum pulling force (6.5) | 0.5 |
| | | **4.0** |

- For identifying the relevant forces, 0.1 points were awarded each for correctly identifying F, W, and N, and 0.2 points for P (0.1 for the component parallel to the plane and 0.1 for the component perpendicular to the plane)

- Each equilibrium of forces (parallel and perpendicular to the plane) was worth 0.5 points.

- When determining $\beta$, 0.1 points were awarded for the correct idea. Correct derivation and the correct result earned 0.3 points. An additional 0.1 points were awarded for verifying that it is indeed a minimum, for example, by using the second derivative.

- If an incorrect $\beta$ led to a consistently incorrect simplified expression for the minimum pulling force, 0.3/0.5 points were awarded.

- Failure to simplify the result for the minimum pulling force resulted in a deduction of 0.1 points.

**Supplemental Part B: Visual and quantitative comparisons of scores assigned to LLM-generated solutions produced six weeks apart, illustrating the phenomenon of temporal variability**

The plots on the left display the score distributions of ten solutions per problem and prompting technique, generated during the initial data collection. In contrast, the plots on the right show the distributions for ten additional solutions per problem and prompting technique, generated approximately six weeks later. Although no backend changes were officially documented by OpenAI for either *GPT-4o* or *o1-preview* during this period, the distributions reveal some noticeable qualitative differences. This suggests potential variability in model behavior over time.

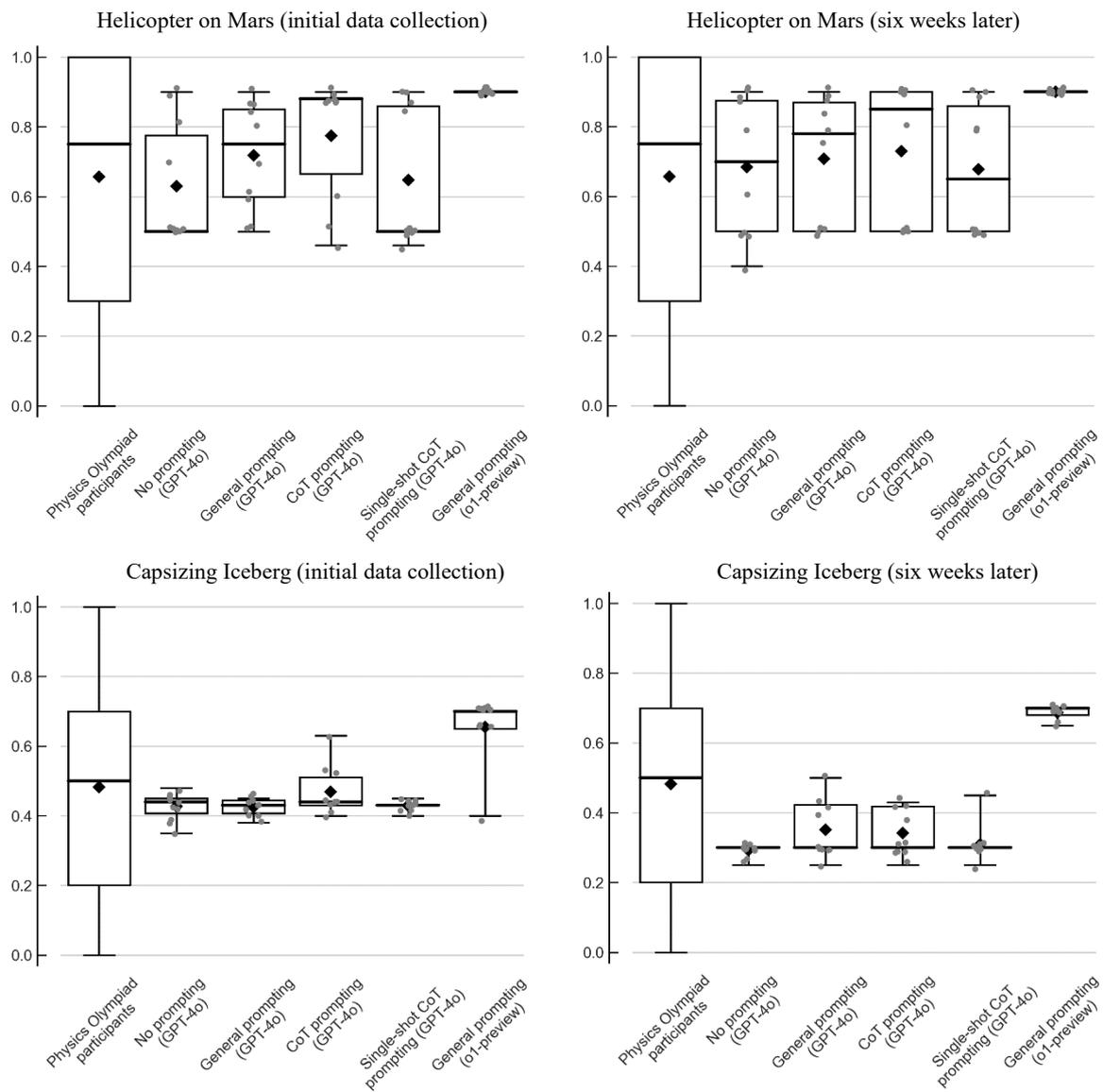

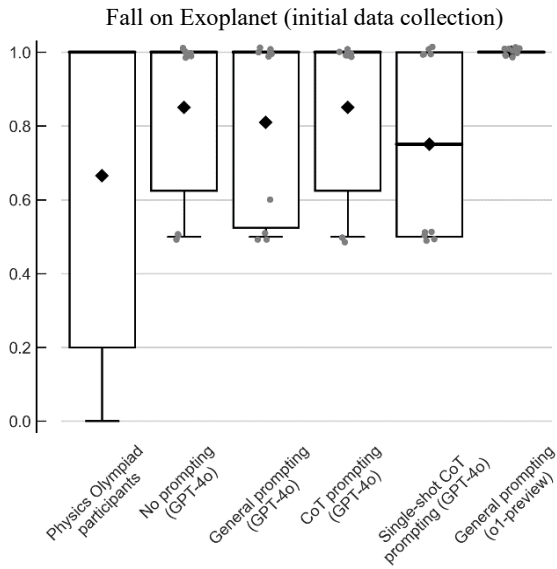
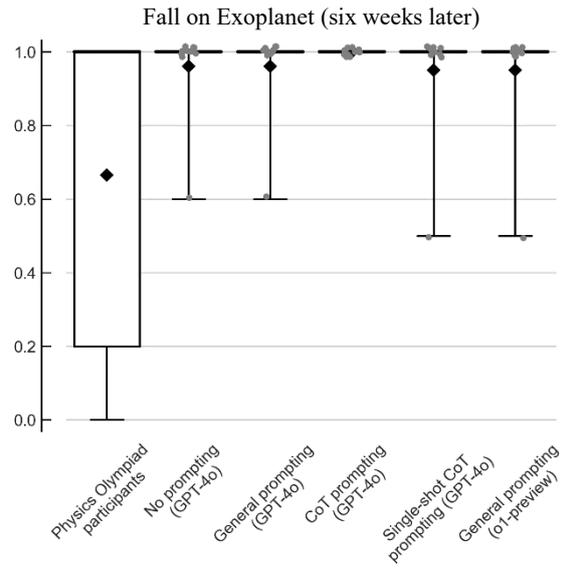
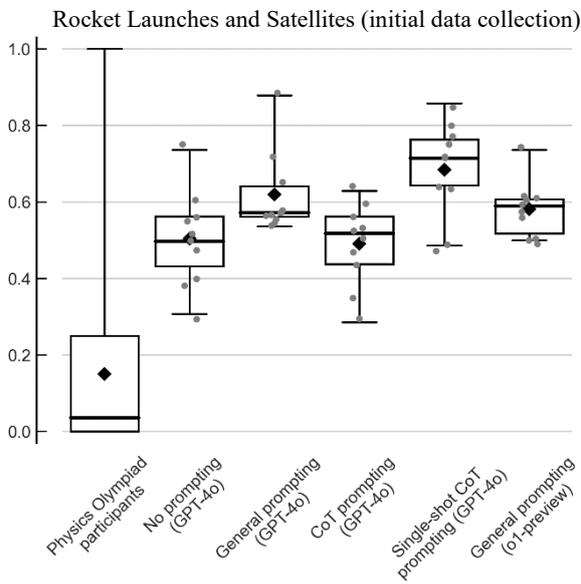
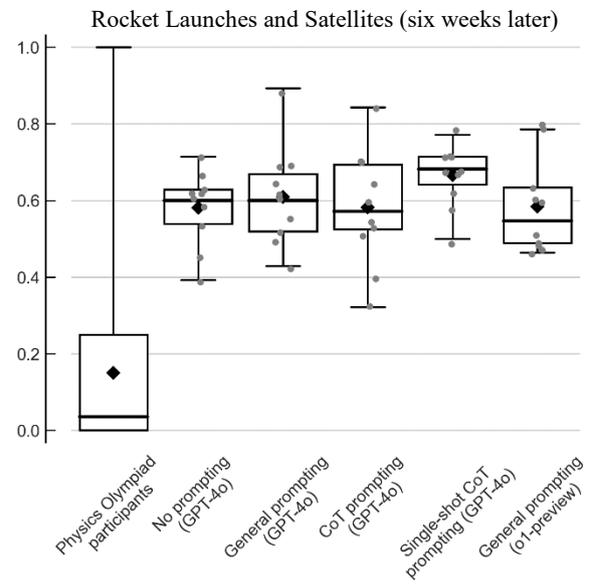
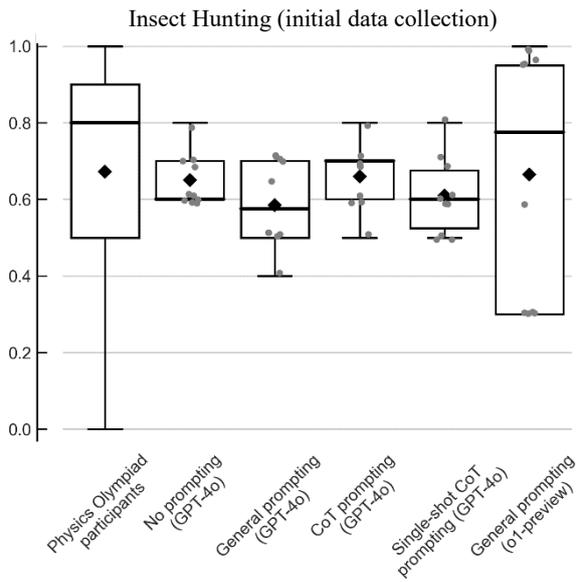
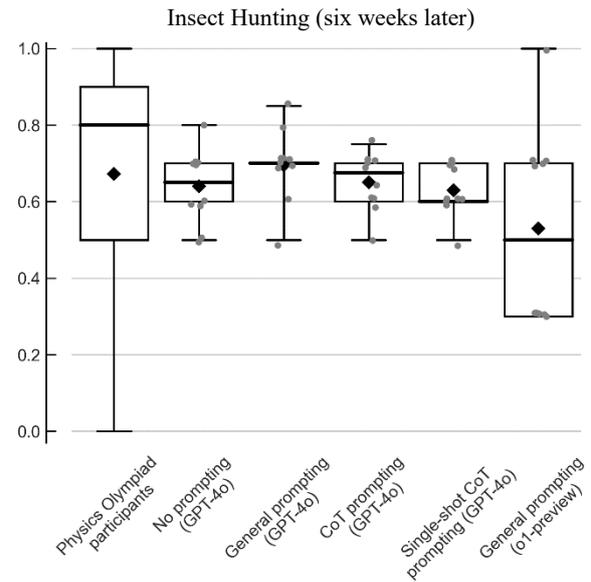

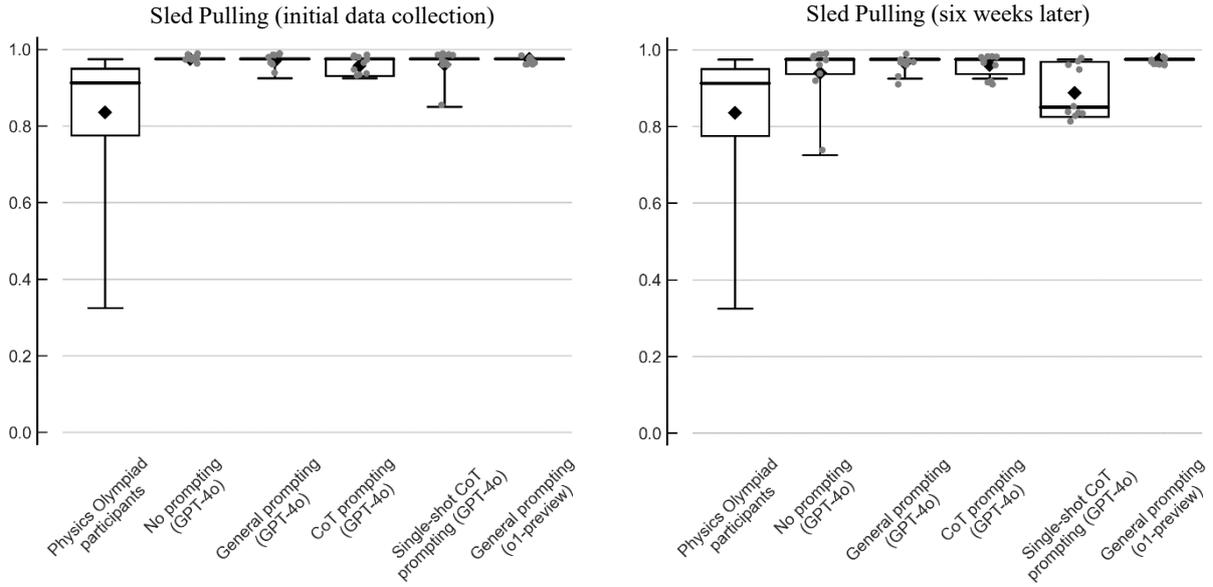

Furthermore, we conducted two-sided Mann–Whitney $U$ tests to quantitatively assess whether the score distributions of LLM-generated solutions differed significantly between the two time points. One comparison was performed for each problem and prompting technique. The computed $p$-values are provided in the following Table:

| Problem | No prompting (*GPT-4o*) | General prompting (*GPT-4o*) | CoT prompting (*GPT-4o*) | Single-shot CoT prompting (*GPT-4o*) | General prompting (*o1-preview*) |
|---|---|---|---|---|---|
| Helicopter on Mars | 0.23 | 0.11 | 0.08 | 0.06 | 0.37 |
| Capsizing Iceberg | 0.63 | 0.91 | 0.78 | 0.68 | 1.00 |
| Fall on Exoplanet | 0.94 | **0.04** | 0.84 | 0.52 | 0.40 |
| Rocket Launches and Satellites | **<0.01** | **0.03** | **0.01** | **<0.01** | 0.61 |
| Insect Hunting | 0.12 | 0.97 | 0.17 | 0.54 | 0.67 |
| Sled Pulling | 0.08 | 0.58 | 0.79 | **0.02** | 1.00 |

*Note.* Statistically significant results ($p < 0.05$) are indicated in bold.

Overall, it is particularly the *Rocket Launches and Satellites* problem where *GPT-4o*-generated solutions differed significantly across the two time points. Although this phenomenon was observed this strong only for this specific problem in our study and thus is unlikely to substantially affect the overall results or conclusions, it highlights an important consideration for future research: When collecting LLM-generated data over an extended period, researchers should be aware that temporal variability in model outputs may arise, potentially influencing the consistency and reliability of the generated output.